\documentclass[11pt]{article}
\usepackage{amsmath,amssymb}

\newcommand{\ba}{\begin{alignat}{3}}
\newcommand{\e}{\epsilon}

\newcommand{\dl}{\delta}
\newcommand{\g}{\gamma}
\newcommand{\gm}{\gamma}

\newcommand{\pa}{\partial}
\newcommand{\tf}{\tfrac}

\newcommand{\mc}{\mathcal}
\newcommand{\D}{\mathcal{D}}

\begin{document}

\begin{flushright}
OU-HET 606
\\
May 14, 2008
\end{flushright}
\vskip1cm
\begin{center}
{\LARGE {\bf Brown-Henneaux's  Canonical Approach to \\[0.2cm] Topologically Massive Gravity}}
\vskip3cm
{\large 
{\bf Kyosuke Hotta,\footnote{hotta@het.phys.sci.osaka-u.ac.jp} 
Yoshifumi Hyakutake,\footnote{hyaku@het.phys.sci.osaka-u.ac.jp} 
\\
Takahiro Kubota\footnote{kubota@het.phys.sci.osaka-u.ac.jp} and 
Hiroaki Tanida\footnote{hiroaki@het.phys.sci.osaka-u.ac.jp}}
}

\vskip1cm
{\it Department of Physics, Graduate School of Science, 
\\
Osaka University, Toyonaka, Osaka 560-0043, Japan}
\end{center}

\vskip1cm
\begin{abstract}

We  analyze the symmetry realized asymptotically on the two dimensional 
boundary of ${\rm AdS}_3$ geometry in topologically 
massive gravity, which consists of the gravitational Chern-Simons term 
as well as the usual Einstein-Hilbert and negative cosmological constant 
terms. Our analysis  is based on the conventional canonical method 
and proceeds along the line completely parallel to  the original 
Brown and Henneaux's. In spite of the presence of the gravitational 
Chern-Simons term, it is confirmed by the canonical method that 
the boundary theory actually has the conformal symmetry satisfying the 
left and right moving Virasoro algebras. The central charges of the 
Virasoro algebras are computed explicitly and are shown to be left-right 
asymmetric due to the gravitational Chern-Simons term. 
It is also argued that the Cardy's formula for the BTZ black hole entropy 
capturing {\it all} higher derivative corrections
agrees with the extended version of 
the Wald's entropy formula. 
The M5-brane system is illustrated as an  application of 
the present calculation. 
\end{abstract}


\vfill\eject




\section{Introduction}\label{Sec1}

The three dimensional spacetime has been one of the interesting testing grounds to uncover
quantum natures of gravity.
Especially, the three dimensional gravity with negative cosmological constant has been paid much attention, 
since this system admits a globally ${\rm AdS}_3$ geometry as a vacuum~\cite{deser}, 
and the black hole solution 
of Ba${\tilde {\rm n}}$ados, Teitelboim and Zanelli (BTZ)  
\cite{btz} as excited states. 
Moreover, this system can equivalently be analyzed by mapping to gauge Chern-Simons action~\cite{witten1}.

One of the interesting properties of the ${\rm AdS}_3$ geometry is that 
there exists an asymptotic symmetry at the boundary, described by  
two dimensional conformal field theory (CFT). By using a canonical 
formalism of the Einstein-Hilbert gravity, Brown and Henneaux 
\cite{brown} successfully 
 constructed left- and right-moving  Virasoro algebras at the boundary, 
 which share  a common  nontrivial value for their  central charges.

The existence of the two dimensional CFT is inferred if we 
embed this system in M-theory~\cite{msw}. 
The low energy limit of the M-theory is well described by eleven 
 dimensional supergravity, and after compactification on Calabi-Yau (CY) 
 3-fold, it becomes five  dimensional supergravity~\cite{CCAF}. 
An M5-brane which wraps on four cycles in CY$_3$ corresponds to 
a string-like black object in five dimensional supergravity, and after 
taking near horizon limit, the geometry becomes  AdS$_3$ $\times$ S$^2$. 
The AdS$_3$ geometry appears after the dimensional reduction of S$^2$ part.
On the other hand, the field theory on the M5-brane is well described by 
two  dimensional CFT after reducing four dimensional part which wraps on four 
 cycles in ${\rm CY}_3$.  In this way we can understand 
 ${\rm AdS}_3/{\rm CFT}_2$ correspondence via the M5-brane wrapping on the 
 ${\rm CY}_3$~\cite{kraus2,kraus1,Krauslec}.
 

The three dimensional theory relevant to the M-theory includes 
both the gravitational Chern-Simons term and other matter fields containing higher derivative terms. 
Let us recall in this connection that Saida and Soda \cite{saida} have 
previously studied the higher derivatives without the Chern-Simons term. 
By applying frame transformation method~\cite{MFF}, they mapped the higher derivative action
to the canonical Einstein-Hilbert one. 
In the case of BTZ black hole, this frame transformation just scales the original metric,
and it becomes possible to calculate the modification of the Virasoro central charges by the 
simple scaling argument. Both left and right central charges scale in the same way and agree with each other,
even if the higher derivative terms are included.

In this paper, we generalize the work of ref.~\cite{saida} by including the gravitational 
Chern-Simons term. This cannot be dealt with by the simple scaling argument, and we need to
consider the canonical formalism in a conventional way.
It has been argued by Gupta and Sen in ref.~\cite{gupta} that the method of 
field redefinition and consistent truncation transforms 
the three dimensional gravity 
theory into the one consisting of only three terms: 
the Einstein-Hilbert, the cosmological constant and the gravitational 
Chern-Simons terms. 
Such a three dimensional theory with negative cosmological constant 
is often referred to  as topologically massive gravity (TMG) \cite{Deser:1982vy,Deser:1981wh}.
The action is given by 
\begin{eqnarray}
& & \mc{S}_{\text{TMG}} = \frac{1}{16 \pi G_N} \int d^3x 
\left ( \mc{L}_{\text{EH}} + \mc{L}_{\text{CS}} \right ), 
\label{eq:action}
\\
& & \mc{L}_{\text{EH}}
=
\sqrt{-G} \Big( R + \frac{2}{\ell^2} \Big), 
\label{eq:EHaction}
\\
& & \mc{L}_{\text{CS}}
=
\frac{\beta}{2} \sqrt{-G} \epsilon^{IJK} \Big( \Gamma^P{}_{IQ} \pa_J 
\Gamma^Q{}_{KP}  + \frac{2}{3} \Gamma^P{}_{IQ} \Gamma^Q{}_{JR} 
\Gamma^R{}_{KP} \Big).
\label{eq:csaction}
\end{eqnarray}
The cosmological constant $-2/ \ell^{2}$ in ${\cal L}_{\rm EH}$ 
is negative and $\beta $ is a coupling constant with the dimension of 
the length. 
The determinant of the three dimensional metric $G_{IJ}$ is denoted by $G$,
the three dimensional Christoffel symbol is by  $\Gamma^P{}_{IQ}$ 
and the capital letters $P$, $I$, $Q$  etc.  label the three dimensional 
space-time indices,  $t$, $r$ and $\phi $. 
Note also that $\epsilon ^{MNO}$ is a covariantly constant tensor 
and $\sqrt{-G}\epsilon ^{MNO}$ is just a constant. 

Notice that (\ref{eq:csaction}) contains third derivative with respect to the time.
The canonical formalism of such a system can be accomplished by using the Ostrogradsky method,
where a new variable is introduced to reduce the number of the time derivative~\cite{ost}.
For the gravitational Chern-Simons term, it is convenient to employ the generalized version of it~\cite{buchbinder,buchbinder2}.
Then it is possible to define the Hamiltonian, and from its variation we can extract the global charges,
such as a mass, an angular momentum and central charges in TMG.
The asymptotic symmetry in  TMG  is 
again described by the left and right moving Virasoro algebras, whose 
central charges are, however,  not the same as we will show later.
The central charges has been derived in previous literatures in several ways~\cite{kraus2,kraus1,sahoo,Solod,gupta}.

The entropy of the black hole may be evaluated by using the Cardy's  
formula together with the modified values of the central charges.  
There is an important remark that the black hole entropy  computed  
on the basis of the Cardy's formula 
should not be compared with the Wald's formula \cite{wald} in its 
original form 
which is applicable only 
for manifestly diffeomorphism invariant theories.  The formula should be 
compared with the one given recently in refs.~\cite{Solod,s}, where a modification 
has been made so that one can include such a term as the Chern-Simons'. 
The agreement of both entropies are confirmed in our canonical formalism.


The structure of our work is as follows. In Sec.~\ref{Sec2}
we present the modified version of the Wald's black hole entropy formula, 
paying a particular attention to higher derivative corrections 
including  those of the Chern-Simons term. The framework of deriving the 
asymptotic symmetry is discussed in Sec.~\ref{Sec3}. The calculation of the 
Virasoro central charges in TMG is given in Sec.~\ref{Sec4}, in the 
Ostrogradsky method 
which is  adapted to the cases of  higher derivative terms. The mass and the 
angular momentum of the BTZ black hole is also discussed,  
including the effects due to the Chern-Simons term. In Sec.~\ref{Sec5}, 
we discuss  all of the higher derivative corrections and compare our black hole entropy 
formulas with the modified version of Wald's  given in Sec.~\ref{Sec2}. 
Our calculation is shown to be useful in the application of M5 system. 
Appendix \ref{App1} is devoted to a detailed discussion on the 
difference between gravitational and Lorentz Chern-Simons terms.
Some of the calculational details are relegated to Appendix \ref{App2}.



\section{Most General Entropy Formula for BTZ Black Holes}
\label{Sec2}

Before starting to discuss the canonical method and the CFT approach 
to the BTZ black hole, we here concentrate on the macroscopic 
treatment  of black hole entropy {\` a} la Wald (See refs.~\cite{p}). We pay a particular 
attention to, and try to include the effects of 
 higher derivative terms together with the Chern-Simons' 
 in  as general a way as possible.
For such a purpose we begin  with the following  Lagrangian 
\begin{equation}
\mc{S}=\frac{1}{16\pi G_N}\int d^3x\sqrt{-G}
\left[f(R_{IJ},G_{IJ})+\frac{2}{\ell_0^2}\right]
+\frac{1}{16\pi G_N}\int d^3x\mathcal{L}_{\text{CS}}.
\label{generalaction}
\end{equation}
Here $f$ is a generally  covariant part and is supposed to contain 
all possible higher derivatives. In three dimensional spacetime, 
the Riemann and  the Ricci tensors have both six components 
and are related by the formula  
 ${R^{IJ}}_{KL}=4{G^{(\,I}}_{(\,K}{R^{J\,)}}_{L\,)}-
 R{G^{(\,I}}_{(\,K}{G^{J\,)}}_{L\,)}$.
Therefore, $f$ is assumed to be a functional of 
the Ricci tensor and the metric only. Note that the negative cosmological 
constant in (\ref{generalaction}) is denoted by $-2/\ell _{0}^{2}$.

The action (\ref{generalaction}) is diffeomorphism invariant up to the total 
derivatives.  Due to the gravitational Chern-Simons term and other higher derivative terms,
the Einstein equation is modified as
\begin{equation}
\frac{1}{2}G^{IJ}\left(f+\frac{2}{\ell_0^2}\right)+
\frac{\partial f}{\partial G_{IJ}}+T^{IJ}
=\beta \epsilon^{KL(I}\mathcal{D}_KR^{J)}_L.
\label{generaleom}
\end{equation}
Here we denote 
\begin{equation}
T^{IJ}=\frac{1}{2}\left(\mathcal{D}_K\mathcal{D}^IP^{KJ}+
\mathcal{D}_K\mathcal{D}^JP^{IK}-
\Box P^{IJ}-G^{IJ}\mathcal{D}_K\mathcal{D}_LP^{KL}\right), \quad
P^{IJ}=\frac{\displaystyle{\partial f}}{\displaystyle{\partial R_{IJ}}}.
\label{eq:tij}
\end{equation}
$\mathcal{D}_I$ is the usual covariant derivative.

Various solutions to (\ref{generaleom}) would  be possible, but it is 
known that the ${\rm AdS}_3$ geometry satisfying
\begin{equation}
\frac{1}{2}G^{IJ}\left(R+\frac{2}{\ell^2}\right)-R^{IJ}=0
\label{ads3sol}
\end{equation}
with some constant $\ell $ 
is certainly  a solution to (\ref{generaleom}). 
This can be seen by the following argument. If (\ref{ads3sol}) is satisfied, 
then the scalar curvature is just a constant 
($R=-6/\ell^{2}$) and the metric and the Ricci 
tensors are proportional   ($R_{IJ}=-2G_{IJ}/\ell^{2}$). 
We can see that (\ref{eq:tij}) and 
the right-hand-side of (\ref{generaleom}) vanish. 
Eq. (\ref{generaleom}) turns out  to be 
 a relation that fixes $\ell $ as a function 
of $\ell _{0}$. This may be regarded as an ``effective'' cosmological constant 
due to the higher derivative terms\footnote{Throughout this paper, 
 the effective cosmological constant $-2/ \ell^2$ always appears in the solutions.}.

The vacuum solution to (\ref{ads3sol}) is the global AdS geometry
with the radius $\ell $
\begin{align}
ds^2 =-\left( 1+\frac{r^2}{\ell ^2}\right)dt^2
+\left(1+\frac{r^2}{\ell ^2}\right)^{-1}dr^2 +r^2 
d\phi ^2 . 
\label{vacuum ads}
\end{align}
As an excited state the  BTZ black hole solution  \cite{btz}
\begin{align}
ds^2&=-N^2dt^2+N^{-2}dr^2+r^2(d\phi+N^\phi dt)^2,\notag\\
N^2=&\left(\frac{r}{\ell}\right)^2+\left(\frac{4G_Nj}{r}\right)^2-8G_Nm
\,\,\,\,\,\,\,\,,\,\,\,\,\,\,\,\,
N^\phi=\frac{4G_Nj}{r^2},
\label{BTZsol}
\end{align}
is also allowed, which preserves the local ${\rm AdS}_3$ symmetry 
and is constructed by global identification of independent points 
on (\ref{vacuum ads}).  In the Einstein-Hilbert gravity with negative 
cosmological constant, 
parameters $m$ and $j$ correspond to the mass and angular momentum 
of the BTZ black hole. In general, however, the definition of the 
mass and angular momentum must be changed by taking into account 
of other higher derivative terms.
As we will see explicitly in Sec.~\ref{Sec4} in the canonical 
formalism,  the effective mass $M$ and the effective
angular momentum $J$ of the BTZ 
black hole (\ref{BTZsol}) are represented by linear combinations 
 of $m$ and $j$  when  the Chern-Simons term is present. 


It is well known that the Bekenstein-Hawking's area law for 
the black hole entropy is modified by Wald's entropy formula
for general covariant theories which include higher derivative 
terms \cite{wald}.
In those  theories, however, which are not manifestly invariant 
under the diffeomorphism, that formula cannot be applied directly 
and must be modified. 
The extended Noether method including the contribution of the 
non-covariant terms was discussed in \cite{s}.
In general, the non-covariant terms such as the gravitational 
Chern-Simons term, which is one of the higher derivative terms, 
modify the Noether charge in a slightly different fashion from 
Wald's formula.  As a result, the black hole entropy receives 
the higher derivative correction further.
In practice we have to determine the corrections to the Wald's 
entropy formula on a case-by-case basis.
For the gravitational Chern-Simons term in  three dimensions  
the additional correction $\Delta S$ to the entropy  is found to be
\begin{equation}
\Delta S=\frac{\beta}{4G_N}\int_H\varepsilon^J_{\,\,\,I}
\Gamma^I_{\,\,\,JK}dx^K.
\end{equation}
Here $\varepsilon^{IJ}$ is a binormal vector on the horizon $H$.
The equivalent results for the three dimensional gravitational 
Chern-Simons term were obtained by several  others 
\cite{Solod,sahoo}.

With the help of this,  the full entropy of the BTZ black hole are, 
therefore, calculated as follows:
\begin{align}
S&=-\frac{1}{8G_N}\oint_{r_+}d\phi\sqrt{G_{\phi\phi}}\frac{\partial f}{\partial R_{IK}}G^{JL}\varepsilon_{IJ}\varepsilon_{KL}
+\frac{\beta}{4G_N}\oint_{r_+}d\phi\,\varepsilon^{JI}\Gamma_{IJ\phi}\notag\\
&=\frac{1}{4G_N}\Omega\oint_{r_+}d\phi\sqrt{G_{\phi\phi}}
+\frac{\beta}{4G_N}\oint_{r_+}d\phi\frac{r_+r_-}{\ell r}
\notag\\
&=\frac{\pi\Omega}{2G_N}r_++\frac{\pi\beta}{2G_N\ell}r_-,
\label{eq:generalentropyformula}
\end{align}
where the conformal factor $\Omega$ is defined by
\begin{equation}
\Omega=\frac{1}{3}G_{IJ}\frac{\partial f}{\partial R_{IJ}}.
\label{conformal}
\end{equation}
This $\Omega $ is just a constant for the BTZ black 
hole solution (\ref{BTZsol}), and $\Omega=1$ for the Einstein-Hilbert action.
By substituting explicit values of $r_\pm=\sqrt{2G_N\ell(\ell m+j)} \pm \sqrt{2G_N\ell(\ell m-j)}$
into (\ref{eq:generalentropyformula}), we finally obtain the entropy formula 
\begin{equation}
S=\frac{\pi}{2G_N}\left\{
\left(\Omega+\frac{\beta}{\ell}\right)
\sqrt{2G_N\ell^2
\left(m+\frac{j}{\ell}\right)}
+\left(\Omega-\frac{\beta}{\ell}\right)
\sqrt{2G_N\ell^2
\left(m-\frac{j}{\ell}\right)}\,\,\right\}.
\label{macroentropy}
\end{equation}
This is the macroscopic entropy including {\it all} higher derivative 
corrections in three dimensions.   In the present  paper we consider 
only the parameter region of $\Omega \ell>\beta>0$ just for simplicity.
The situation where $\Omega<0$ or $\beta>\Omega \ell$ has been  discussed 
in \cite{p}.  In the following sections, by generalizing the original 
Brown-Henneaux's canonical approach, we shall show that  
the expression (\ref{macroentropy}) is in perfect agreement
with the Cardy's formula for the CFT on the two dimensional boundary.



\section{Hamiltonian Formalism and Virasoro Algebras}
\label{Sec3}

As long as the BTZ black holes are concerned, our analyses of the asymptotic symmetry 
associated with (\ref{eq:action}) will go along a line quite 
parallel to those in ref.~\cite{brown} where only the Einstein-Hilbert action is considered. 
It is therefore convenient 
to summarize key ingredients  of Brown and Henneaux's work which are not altered 
when we take the Chern-Simons term into our consideration. 

First of all let us  specify  the boundary conditions so that 
field configurations behave as ``asymptotically AdS$_3$". We require that 
the metric should behave at the spatial infinity $r \to \infty$ as 
\begin{align}
 &G_{tt}=-\frac{r^2}{\ell ^2}+\mathcal{O}(1)
 \,\,\,\,\,\,,\,\,\,\,\,\,
 G_{tr}=\mathcal{O}(r^{-3})
 \,\,\,\,\,\,,\,\,\,\,\,\,
 G_{t\phi}=\mathcal{O}(1), 
 \notag \\
 &G_{rr}=\frac{\ell ^2}{r^2}+\mathcal{O}(r^{-4})
 \,\,\,\,\,\,,\,\,\,\,\,\,
 G_{r\phi}=\mathcal{O}(r^{-3})
 \,\,\,\,\,\,,\,\,\,\,\,\,
 G_{\phi\phi}=r^2+\mathcal{O}(1),  
\label{hypersurface bd}
\end{align}
which is in accordance with the behavior in (\ref{vacuum ads}) and (\ref{BTZsol}). 
The vector fields $({\bar \xi}^{0}, \bar{\xi}^{r}, \bar{\xi }^{\phi}) $ that transform the metric 
while  preserving the boundary conditions (\ref{hypersurface bd}) are not 
strongly restricted but are allowed to be a general class of functions. 
In fact, by using the coordinates of $x^\pm =\frac{\displaystyle{t}}{\displaystyle{\ell}} \pm \phi$, 
the $n$-th Fourier component of the vector fields is given by 
\ba
  \bar{\xi}^t = \frac{\ell}{2} e^{inx^\pm} \Big(1 - \frac{\ell^2 n^2}{2r^2} \Big), \quad
  \bar{\xi}^r = -i \frac{nr}{2} e^{inx^\pm}, \quad
  \bar{\xi}^\phi = \pm \frac{1}{2} e^{inx^\pm} \Big(1 + \frac{\ell^2 n^2}{2r^2} \Big).
  \label{eq:Killing}
\end{alignat}
In this paper we call the above vector fields ``Killing vectors''.
For later use, we assign explicit notations for these Killing vectors:
\begin{align}
 \xi ^{\pm}_{n} \equiv  {\bar \xi}^I \partial_I 
 =  e^{inx^\pm}\left ( \partial_\pm -\frac{\ell ^2
 n^2}{2r^2}\partial_\mp - \frac{inr}{2}\partial_r 
 \right ),
\label{eq:killing2}
\end{align}
where $\partial _{\pm}=\frac{1}{2} (\ell \pa_t \pm 
\pa_\phi )$.
The algebraic structure of the symmetry is encoded in the Killing vector 
and in fact we can directly compute the commutation 
relations of these differential operators
\begin{align}
&\left [ \xi ^{\pm}_{m} ,\xi ^{\pm}_{n} \right]= -i \left(m-n \right)
\xi ^{\pm}_{m+n}, 
\hskip0.5cm 
\left[ \xi ^{+}_{m} , \xi ^{-}_{n} \right] = {\cal O}(r^{-4}). 
\end{align}
This result clearly shows that 
 the  asymptotically ${\rm AdS}_{3}$ spacetime is endowed with the 
 two dimensional conformal symmetry.

In order to evaluate the central extension of the Virasoro algebras, we have 
to know the asymptotic behaviors of the canonical variables and we introduce 
the $(2+1)$-dimensional  decomposition of the three dimensional 
metric $G_{IJ}$  as 
\begin{eqnarray}
G_{IJ}=\left (
\begin{tabular}{cc}
$-N^{2}+N_{k}N^{k}$ & $N_{j}$
\\
$N_{i}$ & $g_{ij}$
\end{tabular}
\right ).
\end{eqnarray}
Here $g_{ij}, (i,j=r , \phi )$ is the two dimensional metric. 
The lapse and shift functions are denoted by 
$N$ and $N_{i}$, respectively. 
The Einstein-Hilbert action is rewritten as usual by 
\begin{eqnarray}
\mc{L}_{\text{EH}}=\sqrt{g}N\left ( r + \frac{2}{\ell ^{2}}+ 
K^{ij}K_{ij}-K^{2} \right ), 
\label{eq:ADMofEH}
\end{eqnarray}
where 
$r$ is the scalar curvature made out of $g_{ij}$, and 
\begin{eqnarray}
K_{ij}&=&\frac{1}{2N}\left ( 
{\dot g}_{ij}-{\D}_{i}N_{j}-{\D}_{j}N_{i}\right ), 
\label{eq:kij}
\\
K&=&g^{ij}K_{ij}. 
\end{eqnarray}
The dot over $g_{ij}$ means the $t$-derivative 
and ${\cal D}_{i}$ is the covariant derivative with respect 
to $g_{ij}$. 
The momentum variable $\pi ^{ij}$ conjugate to $g_{ij}$ is given by 
$\pi ^{ij}=\sqrt{g}(K^{ij}-g^{ij}K)$ for the 
case of Einstein-Hilbert action, and the Hamiltonian 
$\mc{H}$   is the Legendre transform of $\mc{L}_{\text{EH}}$, i.e., 
$\mc{H}=\pi ^{ij}\dot g_{ij} -
\mc{L}_{\text{EH}}$.

The Hamiltonian consists of the usual combination of the constraints 
together with appropriate surface term $Q[\xi]$, 
\begin{gather}
 H[\xi]=\int d^2 x \left ({\xi}^{0} \mathcal{H}+{\xi}^i 
 \mathcal{H}_i \right)+Q[\xi] 
 \label{hamiltonian gene}. 
\end{gather}
The added term $Q[\xi]$ must be determined so that it 
cancels the surface terms produced by the first term 
in (\ref{hamiltonian gene}) under field variation and is a 
generator of the possible surface deformation~\cite{RT}. 
The vector fields $(\xi ^{0}, \xi ^{r}, \xi^{\phi})$ denote such  
allowed surface deformation and are related to the spacetime vector 
$(\bar{\xi }^{0}, \bar{\xi }^{r}, \bar{\xi}^{\phi})$ via
\begin{alignat}{3}
(\xi ^{0}, \xi ^{r}, \xi^{\phi})
 =(N{\bar \xi}^{t}, {\bar \xi}^{r}+N^{r}{\bar \xi}^{t}, 
 {\bar \xi}^{\phi}+N^{\phi}{\bar \xi}^{t}). 
 \label{eq:localcoordinate}
\end{alignat}
The asymptotic behaviors (\ref{hypersurface bd}) are now translated into 
those of the canonical variables  as 
\begin{align}
&g_{rr}=\frac{\ell ^2}{r^2}+\mathcal{O}(r^{-4})
\,\,\,\,\,\,,\,\,\,\,\,\,
g_{r\phi}=\mathcal{O}(r^{-3})
\,\,\,\,\,\,,\,\,\,\,\,\,
g_{\phi\phi}=r^2 +\mathcal{O}(1),
\label{2d-g-bd}
\\
 &N=\frac{r}{\ell}+\mathcal{O}(r^{-1})
 \,\,\,\,\,\,,\,\,\,\,\,\,
 N^{r}=\mathcal{O}(r^{-1})
 \,\,\,\,\,\,,\,\,\,\,\,\,
 N^\phi=\mathcal{O}(r^{-2}).
 \label{r-s-bd}
 \end{align}
The behaviors of the canonical conjugate variables are also derived with the 
help of (\ref{eq:kij}), (\ref{2d-g-bd}) and (\ref{r-s-bd}):
\begin{align}
 &\pi^{rr}=\mathcal{O}(r^{-1})
 \,\,\,\,\,\,,\,\,\,\,\,\,
 \pi^{r\phi}=\mathcal{O}(r^{-2})
 \,\,\,\,\,\,,\,\,\,\,\,\,
 \pi^{\phi\phi}=\mathcal{O}(r^{-5}) .
 \label{conju-bd}
\end{align}
It has been known that  conditions (\ref{2d-g-bd}), 
(\ref{r-s-bd}) and (\ref{conju-bd})
are preserved under the Hamiltonian evolution provided that 
we impose the Hamiltonian constraints. The generator $Q[\xi]$ 
in (\ref{hamiltonian gene}) is found by taking into account  the asymptotic 
behaviors of the canonical variables up to a constant term, which is 
adjusted so that the charge $Q[\xi]$ vanishes for the globally AdS space. 

The algebraic structure of symmetric transformation group is given by the
Poisson bracket algebra of Hamiltonian generator $H[\xi]$:
\begin{align}
 \left\{H[\xi ],H[\eta
 ]\right\}_{\text{P}}=H\big[[\xi,\eta]\big]+K[\xi,\eta] , 
 \label{p-b}
\end{align}
where $K[\xi,\eta]$ is the possible central extension. 
The central charge may be evaluated by noting that 
the Dirac bracket $\left \{ Q[\xi ], Q[\eta ] \right \}_{\text{D}}$
is the change of $Q[\xi]$ under the surface deformation 
due to $Q[\eta]$, i.e., $\delta_\eta Q[\xi]=\left\{Q[\xi ],Q[\eta ]\right\}_\text{D}$.
The charge $Q[\xi]$ forms a conformal group with a central extension 
$\left \{ Q[\xi ], Q[\eta] \right \}_{\text{D}}
=Q \big[ [ \xi,\eta ] \big ] + K[\xi,\eta] $, and we immediately get
$\delta _{\eta} Q[\xi] =Q\big[[\xi,\eta]\big]+K[\xi,\eta]$.
Since $Q\big[[\xi,\eta]\big]=0$ if we set the initial condition so that 
the charge vanishes for a globally AdS space,  the evaluation 
of the central charge reduces to 
\begin{align}
 K\big[ \xi,\eta \big]=\delta_\eta Q[\xi]. 
 \label{central-ext}
\end{align}
In the case of Einstein-Hilbert action, the explicit form is given by
\ba
&\delta _{\eta}Q[\xi] \notag
\\
&= \int d\phi \left [ \sqrt{g} S^{ijkr}\{ \xi ^{0}{\cal D}_{k}\delta _{\eta} g_{ij} - 
{\cal D}_{k} \xi ^{0} \delta _{\eta} g_{ij}
\}+2\xi ^{i}\pi ^{jr} \delta _{\eta}g_{ij}
+2\xi _{i}\delta _{\eta}{\pi ^{ir}} -\xi ^{r}\pi ^{ij}\delta _{\eta}g_{ij}
\right ], \label{eq:dQ}
\end{alignat}
where $S^{ijkl}$ is defined by
\begin{eqnarray}
S^{ijkl}=\frac{1}{2}\left (
g^{ik}g^{jl}+g^{il}g^{jk}-2g^{ij}g^{kl}
\right ).
\label{eq:sijkl}
\end{eqnarray}
(Derivation of the above equations will be explained in the case of TMG in Sec.~\ref{Sec4}.)

Putting the Killing vectors  (\ref{eq:killing2}) for $\xi $, 
we define the Virasoro generators  by 
$\hat L_{n}^{\pm}=Q[\xi ^{\pm}_{n}]$. 
Replacing the Dirac brackets by a commutator 
($\{ , \}_{\rm D} \to -i [ , ] $), 
the  commutation relations become
\begin{eqnarray}
& & [ \hat L_{m}^{+}, \hat L_{n}^{+}]=(m-n)\hat L_{m+n}^{+}+\frac{c_{L}}{12}
m(m^{2}-1)\delta _{m+n, 0}, 
\nonumber
\\
& & [ \hat L_{m}^{-}, \hat L_{n}^{-}]=(m-n)\hat L_{m+n}^{-}+\frac{c_{R}}{12}
m(m^{2}-1)\delta _{m+n, 0}, 
\nonumber \\
& & [L_{m}^{+}, L_{n}^{-}]=0,
\end{eqnarray}
and the central charges have been calculated in \cite{brown} as
\begin{eqnarray}
c_{L}=c_{R}=\frac{3\ell }{2G_{N}}.
\end{eqnarray}
Once we get the central charges, it is straightforward to obtain 
the BTZ black hole entropy by using the Cardy's formula
\begin{alignat}{3}
 S &= 2\pi \sqrt{\frac{1}{6}c_{L}L_{0}^{+}}+2\pi\sqrt{\frac{1}{6}c_{R}L_{0}^{-}} \notag
 \\
 &= \frac{\pi}{2G_{N}}\sqrt{2G_{N}
 \ell ^{2}\left(  m + \frac{j}{\ell} \right ) }+
 \frac{\pi}{2G_{N}}\sqrt{2G_{N}\ell ^{2} \left(  m - \frac{j}{\ell} \right)} . 
 \label{cardy}
\end{alignat}
Here $L_{0}^{\pm}$ are the eigenvalues of $\hat L_{0}^{\pm }$ and 
are related to the mass $m$  and angular momentum $j$ of the black 
hole by the formulae
$L_{0}^{+}+L_{0}^{-}=m\ell$ and  $L_{0}^{+}-L_{0}^{-}=j$.

In Sec.~\ref{Sec2} we have seen that even in the presence of 
higher derivative interactions, (\ref{vacuum ads}) and 
(\ref{BTZsol}) are still solutions to the equations of motion 
with the effective cosmological constant $-2/ \ell^2$. There occurs a phenomenon 
of rescaling of the charges due to the higher derivative 
terms, and in fact the first term in the entropy formula 
(\ref{eq:generalentropyformula}) is rescaled by the factor $\Omega $ defined 
by (\ref{conformal}). The question that we would like to address 
ourselves here is 
what this rescaling phenomenon looks like in the CFT framework~\cite{saida}. 

Let us  start with the diffeomorphism invariant Lagrangian without the gravitational Chern-Simons term,
\begin{align}
\mc{L}= \sqrt{-G}\left [ f(R_{IJ},G_{IJ})+\frac{2}{\ell_0^2}\right ]. \label{high-gravity}
\end{align}
Higher derivative terms are again included in  $f$. 
The important point is that the Lagrangian constructed out of the metric and the Ricci tensor is 
equivalent to the Einstein-Hilbert Lagrangian with matter fields after the frame transformation~\cite{MFF}.
The metric $\tilde{G}^{IJ}$ in the Einstein frame is defined to be
\ba
  \tilde{G}^{IJ} &= \Big| \det \Big( \frac{\pa \mc{L}}{\pa R_{KL}} \Big) \Big|^{-1} 
  \frac{\pa \mc{L}}{\pa R_{IJ}},
\end{alignat}
and when BTZ solution (\ref{BTZsol}) is substituted into the above expression, we obtain
\ba
  \tilde{G}_{IJ} &= \Omega^2 G_{IJ}.
\end{alignat}
The conformal factor $\Omega$ is already defined in  (\ref{conformal}) and becomes a constant here.
This means that canonical variables are scaled like $\tilde{g}_{ij} = \Omega^2 g_{ij}$,
$\tilde{N} = \Omega N$, $\tilde{N}^i = N^i$ and $\tilde{\pi}^{ij} = \Omega^{-1} \pi^{ij}$.
From these scaling rules, we find that the mass $M$, the angular momentum $J$ and the central charges $c_L$ and $c_R$,
which are evaluated by using  (\ref{eq:dQ}), are multiplied by $\Omega$ as
\ba
  M = \Omega m, \qquad J = \Omega j , \qquad c_L = c_R = \Omega \frac{3 \ell}{2 G_N}. \label{eq:massang}
\end{alignat}
Then the eigenvalues of the Virasoro generators are also linearly scaled as
$L_{0}^{\pm} = \frac{1}{2} ( M\ell \pm J ) = \frac{1}{2} \Omega ( m\ell \pm j )$,
and these considerations lead us to conclude that 
effects of higher derivative terms to the BTZ black hole 
are all summarized by  the rescaling i.e., 
\begin{gather}
 S=\frac{\pi}{2G_{N}} \Omega \sqrt{2G_{N}
 \ell ^{2}\left(  m + \frac{j}{\ell} \right ) }+
 \frac{\pi}{2G_{N}} \Omega \sqrt{2G_{N}\ell ^{2} \left(  m - \frac{j}{\ell} \right)} . 
\end{gather}
Of course this agrees with (\ref{macroentropy}) with $\beta =0~$\cite{saida}. 
We will come to this scaling rule again in Sec.~\ref{Sec5} 
after establishing the canonical formalism and CFT description 
of the Chern-Simons term.




\section{Generalization to Topologically Massive Gravity}
\label{Sec4}

\subsection{Canonical Formalism of Charges in TMG}

In this section we investigate the canonical formalism
of TMG  with negative cosmological constant, and derive the 
expression of global charges in this system. 
From these global charges, we will confirm that the mass 
and the angular momentum of the BTZ black hole and the central charges 
of the boundary CFT are all modified in TMG. 


Let us apply the ADM decomposition to the Lagrangian of the 
topologically massive gravity. 
The action of TMG  is given by (\ref{eq:action}), and the Einstein-Hilbert term with 
negative cosmological constant is decomposed as in (\ref{eq:ADMofEH}).  
After straightforward but tedious calculation, 
 the gravitational  Chern-Simons term can be decomposed up to total 
 derivative  terms into~\cite{Deser:1991qk,buchbinder2}
\ba
  &\sqrt{-G} \epsilon^{IJK} \Big( \Gamma^P{}_{IQ} \pa_J \Gamma^Q{}_{KP}
  + \frac{2}{3} \Gamma^P{}_{IQ} \Gamma^Q{}_{JR} \Gamma^R{}_{KP} \Big) \notag
  \\
  &\cong \sqrt{g} \e^{mn} \Big\{ 2 \dot K_{mk} K_n{}^k + \dot \gamma^x{}_{my} \gamma^y{}_{nx}
  - 2 \pa_k N \D_n K_m{}^k + 2 \D_n \pa_k N K_m{}^k  \notag
  \\
  &\qquad\qquad\quad
  + N K_y{}^x \partial_m \gamma^y{}_{nx} - \partial_m N K_y{}^x \gamma^y{}_{nx}
  - N \D_m K_y{}^x \gamma^y{}_{nx} \notag
  \\
  &\qquad\qquad\quad
  - 2 N^i K_i{}^l \D_n K_{ml} + 2 \D_n N^i K_i{}^l K_{ml} + 2 N^i \D_n K_i{}^l K_{ml} \notag
  \\
  &\qquad\qquad\quad
  + 2 \D_k N^i K_{ni} K_m{}^k + \D_j N^i \partial_m \gamma^j{}_{ni}
  - \D_m \D_j N^i \gamma^j{}_{ni} \Big\} \notag
  \\
  &\cong 2 \sqrt{g} \e^{mn} \dot K_{mk} K_n{}^k 
  + \sqrt{g} N \big\{ 4 \e^{mn} \D_k \D_n K_m{}^k - 2 A^{kl}K_{kl} \big\} \notag
  \\
  &\quad\,
  + \sqrt{g} N^i \big\{ - 4 \e^{mn} K_i{}^l \D_n K_{ml} - 2 \e^{mn} \D_k(K_{ni}K_m{}^k)
  + \e_{ij} \pa^j r + 2 \D_k A_i{}^k \big\}. \label{eq:ADMofGCS}
\end{alignat}
Here $\e^{mn}$ is a covariantly constant antisymmetric tensor in two dimensions, and $\D_i$ 
is the covariant derivative. In the above, we used that the Riemann tensor in two dimensions is expressed 
by the scalar curvature as $r^i{}_{jmn}=\frac{1}{2}(\dl^i_m g_{jn}-\dl^i_n g_{jm})r$.
$A^{ij}$ is a symmetric tensor which is defined by the following equation.
\ba
  &\int d^3x \sqrt{g}  \e^{mp} \dot \g^l{}_{mn} \g^n{}_{pl} 
  = - \int d^3x \sqrt{g}  A^{ij} \dot g_{ij}. \label{eq:defA}
\end{alignat}
The dot is used to represent time derivative $\pa_t$.
By defining
\ba
  T^{ijk}_{mno} &\equiv \tf{1}{2} ( \dl^k_m \dl^{(i}_o \dl^{j)}_n 
  + \dl^k_n \dl^{(i}_o \dl^{j)}_m - \dl^k_o \dl^{(i}_m \dl^{j)}_n ), 
  \label{eq:tijkmno}
\end{alignat}
the time derivative of the affine connection is expressed as $\dot \g^l{}_{mn} = g^{lo} T^{ijk}_{mno} \D_k \dot g_{ij}$.
After the partial integration in  (\ref{eq:defA}), $A^{ij}$ is explicitly written as
\ba
  A^{ij} &= \e^{mp} g^{lo} T^{ijk}_{mno} \D_k \g^n{}_{pl} \notag
  \\
  &= \tf{1}{4} \e^{kl} \D_k \g^i{}_{l}{}^j + \tf{1}{4} \e^{il} \D_k \g^k{}_{l}{}^j 
  - \tf{1}{4} \e^{il} \D_k \g^j{}_{l}{}^k + \big(i \leftrightarrow j \big). 
\label{eq:aij}
\end{alignat}
Since $A^{ij}$ depends on the affine connection in an explicit way, it does not behave as a tensor.
The derivation of  (\ref{eq:ADMofGCS}) is explained in Appendix 
\ref{App1} by focusing on the difference from  Lorentz Chern-Simons term\footnote{The notations employed
here are slightly different from those in ref.~\cite{Deser:1991qk,buchbinder2}.}.

The explicit form of the extrinsic curvature is given by (\ref{eq:kij}).
Then  (\ref{eq:ADMofGCS}) contains third derivatives with respect to time. 
It is known that the canonical formalism
of such system is done by using Ostrogradsky method \cite{ost} 
in which Lagrange multiplier is introduced.
For instance, if there is a Lagrangian $\mc{L}(g,\dot g, \ddot g)$,
then we define $\mc{L}^*(g,\dot g,h,\dot h,v) = \mc{L}(g,h,\dot h) 
+ v(\dot g -h)$ and
construct the Hamiltonian in the usual way. In the case of TMG , 
it is useful to apply modified version of Ostrogradsky method as 
discussed in ref.~\cite{buchbinder,buchbinder2}.
In the modified Ostrogradsky method, the extrinsic curvature is 
dealt with an independent variable.
At the same time, Lagrange multiplier should be introduced to give 
a proper constraint.  Following this prescription, the Lagrangian of 
the TMG   is given by
\ba
  \mc{L}_\text{TMG} &= \mc{L}_\text{EH} + \mc{L}_\text{CS} \notag
  \\
  &= \sqrt{g} N \Big( r + \frac{2}{\ell^2} + K^{ij}K_{ij} - K^2 \Big)
  + v^{ij} (\dot g_{ij} - 2NK_{ij} - 2 \D_i N_j) \notag
  \\
  &\quad\,
  + \beta \sqrt{g} \e^{mn} \dot K_{mk} K_n{}^k 
  + \beta \sqrt{g} N \Big( 2 \e^{mn} \D_k \D_n K_m{}^k - A^{kl}K_{kl} \Big) \notag
  \\
  &\quad\,
  + \beta \sqrt{g} N^i \Big\{ - 2 \e^{mn} K_i{}^l \D_n K_{ml} - \e^{mn} \D_k(K_{ni}K_m{}^k)
  + \frac{1}{2} \e_{ij} \pa^j r + \D_k A_i{}^k \Big\}.
\end{alignat}
Canonical variables in this Lagrangian are $g_{ij}$, $\dot g_{ij}$, $K_{ij}$ and $\dot K_{ij}$, and
$N$, $N^i$ and $v^{ij}$ are Lagrange multipliers. 
Note that $v_{ij}$, which is not a tensor, is symmetric under the exchange of indices.

By using this Lagrangian, we can construct the Hamiltonian in the 
canonical procedure.  As usual, momenta conjugate to 
$\dot g_{ij}$ and $\dot K_{ij}$ are defined as
\ba
  \pi^{ij} &\equiv \frac{\dl \mc{L}_\text{TMG}}{\dl \dot g_{ij}} = v^{ij}, \notag
  \\
  \Pi^{ij} &\equiv \frac{\dl \mc{L}_\text{TMG}}{\dl \dot K_{ij}} = \beta \sqrt{g} \e^{ik} K_k{}^j.
\end{alignat}
Note that not $\pi^{ij}$ and $\Pi^{ij}$ but $g^{-\frac{1}{2}} \pi^{ij}$ and $g^{-\frac{1}{2}} \Pi^{ij}$ 
do behave like tensors.
From the second equation, we see that $\Pi_{ij}$ and $K_{ij}$ are not 
independent and the system is constrained.
Again, such a kind of the constraint should be taken into account 
 by introducing Lagrange multiplier in the Hamiltonian formalism.
Up to total derivative terms, the Hamiltonian of TMG  is expressed as
\ba
  &\mc{H}_\text{TMG} \notag
  \\
  &= \pi^{ij} \dot g_{ij} + \Pi^{ij} \dot K_{ij} - \mc{L}_\text{TMG} 
  + f_{ij} \big( \Pi^{ij} - \beta \sqrt{g} \e^{ik} K_k{}^j \big) \notag
  \\
  &\cong \sqrt{g} N \Big\{ - r - \frac{2}{\ell^2} - K^{kl}K_{kl} + K^2 
  - 2\beta \e^{mn} \D_k \D_n K_m{}^k + \big(2 g^{-\frac{1}{2}} \pi^{kl} + \beta A^{kl} \big) 
  K_{kl} \Big\} \notag
  \\
  &\quad\,
  + \sqrt{g} N^i \Big\{ 2\beta \e^{mn} K_i{}^l \D_n K_{ml} + \beta \e^{mn} \D_k(K_{ni}K_m{}^k)
  - \frac{1}{2} \beta \e_{ij} \pa^j r - \D_j \big(2 g^{-\frac{1}{2}} \pi_i{}^j + \beta A_i{}^j \big) 
  \Big\} \notag
  \\[0.1cm]
  &\quad\,
  + f_{ij} \big( \Pi^{ij} - \beta \sqrt{g} \e^{ik} K_k{}^j \big),
\end{alignat}
where $f_{ij}$ is the Lagrange multiplier.
The validity of this Hamiltonian will be confirmed by explicitly deriving 
the equations of motion in TMG.
In fact we will show below that a part of the equations of motion 
matches with the one obtained
from the Lagrangian formalism in three dimensions.

Let us consider the variation of the Hamiltonian by fluctuating 
$g_{ij}$, $\pi^{ij}$,
$K_{ij}$, $\Pi^{ij}$, $N$, $N^i$ and $f_{ij}$. The variation of the 
Hamiltonian contains
total derivative terms. Though those terms are very important to 
define charges, we neglect them 
for a while to make the argument as simple as possible. Then up to 
total derivative terms, the variation of the
Hamiltonian under the fluctuations of $N$, $N^i$ and $f_{ij}$ is calculated as
\ba
  &\dl_{(N,N^i,f_{ij})} \mc{H}_\text{TMG} \notag
  \\
  &= \dl N \sqrt{g} \Big\{ - r - \frac{2}{\ell^2} - K^{kl}K_{kl} + K^2 
  - 2\beta \e^{mn} \D_k \D_n K_m{}^k + \big( 2 g^{-\frac{1}{2}} \pi^{kl} + \beta A^{kl} \big) 
  K_{kl} \Big\} \notag
  \\
  &\quad\,
  + \dl N^i \sqrt{g} \Big\{ 2\beta \e^{mn} K_i{}^l \D_n K_{ml} + \beta \e^{mn} \D_k(K_{ni}K_m{}^k)
  - \frac{1}{2} \beta \e_{ij} \pa^j r 
  - \D_j \big(2 g^{-\frac{1}{2}} \pi_i{}^j + \beta A_i{}^j \big) \Big\} \notag
  \\
  &\quad\,
  + \dl f_{ij} \Big\{ \Pi^{ij} - \beta \sqrt{g} \e^{ik} K_k{}^j \Big\}. \label{eq:vH1}
\end{alignat}
From this we see that the canonical variables are constrained like
\ba
  &- r - \frac{2}{\ell^2} - K^{kl}K_{kl} + K^2 
  - 2\beta \e^{mn} \D_k \D_n K_m{}^k + (2 g^{-\frac{1}{2}} \pi^{kl} + \beta A^{kl}) K_{kl} = 0, \label{eq:piplusA}
  \\
  &2\beta \e^{mn} K_i{}^l \D_n K_{ml} + \beta \e^{mn} \D_k(K_{ni}K_m{}^k)
  - \frac{1}{2} \beta \e_{ij} \pa^j r 
  - \D_j \big(2 g^{-\frac{1}{2}} \pi_i{}^j + \beta A_i{}^j \big) = 0,
  \\
  &\Pi^{ij} - \beta \sqrt{g} \e^{ik} K_k{}^j = 0. \label{eq:Pi}
\end{alignat}
Note that neither $g^{-\frac{1}{2}} \pi^{ij}$ nor $A^{ij}$ behaves 
like tensors. The linear combination
of $(2g^{-\frac{1}{2}} \pi^{ij} + \beta A^{ij})$, however, does behave as a tensor.

Next, up to total derivative terms, the variation of the
Hamiltonian under the fluctuations of $g_{ij}$, $K_{ij}$, $\pi^{ij}$ and $\Pi^{ij}$ is calculated as
\ba
  &\dl_{(g_{ij},K_{ij},\pi^{ij},\Pi^{ij})} \mc{H}_\text{TMG} \notag
  \\
  &= \dl \pi^{ij} \Big\{ 2 N K_{ij} + 2 \D_i N_j \Big\} 
  + \dl \Pi^{ij} \Big\{ f_{ij} \Big\} 
  + \dl (\sqrt{g} \beta A^{ij}) \Big\{ NK_{ij} + \D_i N_j \Big\} \notag
  \\
  &\quad\,
  + \dl g_{ij} \sqrt{g} \Big\{ N \Big( r^{ij} - \frac{1}{2}g^{ij}r - \frac{1}{\ell^2}g^{ij} \Big) 
  + 2 N (K^{ik}K^j{}_k - K K^{ij}) - \frac{1}{2} N g^{ij} (K^{kl} K_{kl} - K^2) \notag
  \\
  &\qquad\quad
  - (\mc{D}^i \mc{D}^j N - g^{ij} \mc{D}_k \mc{D}^k N) 
  + (2 g^{-\frac{1}{2}} \pi^{k(i} + \beta A^{k(i}) \mc{D}_k N^{j)} 
  - \frac{1}{2} \D_k \big( N^k (2 g^{-\frac{1}{2}} \pi^{ij} + \beta A^{ij}) \big) \notag
  \\
  &\qquad\quad
  + 2 \beta \e^{mn} N \D^i \D_n K_m{}^j 
  - 2 \beta \e^{mn} N^k K_k{}^i \D_n K_m{}^j 
  + 2 \beta \e^{mn} g^{kl} T^{ijz}_{npl} \D_z (N^o K_o{}^p K_{mk}) \notag
  \\[0.1cm]
  &\qquad\quad
  - 2 \beta \e^{mn} g^{kl} g^{op} \D_z \big( - \D_k N K_{mo} T^{ijz}_{nlp}
  + N \D_o K_{ml} T^{ijz}_{knp} + 2 N \D_n K_{o(l} T^{ijz}_{m)kp} \big) \notag
  \\[0.1cm]
  &\qquad\quad
  - \beta \e^{mn} N^k \D^i(K_{nk}K_m{}^j) 
  + 2 \beta \e^{mn} g^{qk} g^{lo} \D_z \big( N^p K_{mk} K_{l(p} T^{ijz}_{n)qo} 
  + N^p K_{np} K_{l(k} T^{ijz}_{m)qo} \big) \notag
  \\
  &\qquad\quad
  - \frac{1}{2} \beta \e^{mn} \dl^i_m N^j \pa_n r 
  - \frac{1}{2} \beta \e^{mn}  \D_n N_m r^{ij} 
  + \frac{1}{2} \beta \e^{mn} S^{ijkl} \D_k \D_l \D_n N_m \Big\} \label{eq:vH2}
  \\
  &\quad\,
  + \dl K_{ij} \sqrt{g} \Big\{ -2 N K^{ij} + 2 N g^{ij} K - 2\beta \e^{ik} \D_k \D^j N 
  + N \big( 2 g^{-\frac{1}{2}} \pi^{ij} + \beta A^{ij} \big) + 2\beta \e^{mn} N^i \D_n K_m{}^j \notag
  \\
  &\qquad\quad
  - 2\beta \e^{ik} \D_k (N^l K_l{}^j) + \beta \e^{ik} \D_l N^j K_k{}^l - 
  \beta \e^{ik} \D^j N^l K_{lk}
  + \beta \e^{ik} f_k{}^j \Big\}. \notag
\end{alignat}
Here 
$S^{ijkl}$ has been defined by (\ref{eq:sijkl}). 
Note that $A^{kl}$ is a function of $g_{ij}$ and $\dl A^{kl}$ depends 
linearly  on $\dl g_{ij}$.
From this, equations of motion for canonical variables are written as
\ba
  \dot g_{ij} &= 2NK_{ij} + 2 \D_{(i} N_{j)}, \label{eq:g}
  \\
  \dot K_{ij} &= f_{ij}, 
  \\
  \dot \pi^{ij} &= - \sqrt{g} \Big\{ N \Big( r^{ij} - \frac{1}{2}g^{ij}r - \frac{1}{\ell^2}g^{ij} \Big) 
  + 2 N (K^{ik}K^j{}_k - K K^{ij}) - \frac{1}{2} N g^{ij} (K^{kl} K_{kl} - K^2) \notag
  \\
  &\quad\,
  - (\mc{D}^i \mc{D}^j N - g^{ij} \mc{D}_k \mc{D}^k N) 
  + (2 g^{-\frac{1}{2}} \pi^{k(i} + \beta A^{k(i}) \mc{D}_k N^{j)} 
  - \frac{1}{2} \D_k \big( N^k (2 g^{-\frac{1}{2}} \pi^{ij} + \beta A^{ij}) \big) \notag
  \\
  &\quad\,
  + 2 \beta \e^{mn} N \D^i \D_n K_m{}^j 
  - 2 \beta \e^{mn} N^k K_k{}^i \D_n K_m{}^j 
  + 2 \beta \e^{mn} g^{kl} T^{ijz}_{npl} \D_z (N^o K_o{}^p K_{mk}) \notag
  \\[0.1cm]
  &\quad\,
  - 2 \beta \e^{mn} g^{kl} g^{op} \D_z \big( - \D_k N K_{mo} T^{ijz}_{nlp}
  + N \D_o K_{ml} T^{ijz}_{knp} + 2 N \D_n K_{o(l} T^{ijz}_{m)kp} \big) \notag
  \\[0.1cm]
  &\quad\,
  - \beta \e^{mn} N^k \D^i(K_{nk}K_m{}^j) 
  + 2 \beta \e^{mn} g^{qk} g^{lo} \D_z \big( N^p K_{mk} K_{l(p} T^{ijz}_{n)qo} 
  + N^p K_{np} K_{l(k} T^{ijz}_{m)qo} \big) \notag
  \\
  &\quad\,
  - \frac{1}{2} \beta \e^{mn} \dl^i_m N^j \pa_n r 
  - \frac{1}{2} \beta e \e^{mn}  \D_n N_m r^{ij} 
  + \frac{1}{2} \beta \e^{mn} S^{ijkl} \D_k \D_l \D_n N_m \label{eq:dpi}
  \\
  &\quad\,
  - \frac{1}{2} \beta \e^{mn} T^{xyz}_{mlo} \Big( \dot g_{xy} g^{oi}g^{pj} \D_z \g^l{}_{np} 
  + g^{op} g^{lr} \dot g_{xy} \D_k \g^q{}_{np} T^{ijk}_{zqr} 
  - 2 g^{op} g^{qr} \dot g_{xy} \D_k \g^l{}_{q(p} T^{ijk}_{n)zr} \notag
  \\
  &\qquad\qquad
  - g^{op} g^{lq} \D_k \D_z \dot g_{xy} T^{ijk}_{npq} 
  + g^{op} g^{lr} \D_k \dot g_{xy} \g^q{}_{np} T^{ijk}_{zqr} 
  - 2 g^{op} g^{qr} \D_k \dot g_{xy} \g^l{}_{q(p} T^{ijk}_{n)zr} \Big) \Big\} , \notag
  \\
  \dot \Pi^{ij} &= - \sqrt{g} \Big\{ - 2N K^{ij} + 2 N g^{ij} K - 2\beta \e^{ik} \D_k \D^j N 
  + N \big(2 g^{-\frac{1}{2}} \pi^{ij} + \beta A^{ij} \big) + 2 \beta \e^{mn} N^i \D_n K_m{}^j \notag
  \\
  &\quad\,
  - 2\beta \e^{ik} \D_k (N^l K_l{}^j) + \beta \e^{ik} \D_l N^j K_k{}^l 
  - \beta \e^{ik} \D^j N^l K_{lk} + \beta \e^{ik} f_k{}^j \Big\}. \label{eq:dotPi}
\end{alignat}
Note that (\ref{eq:g}) is used in (\ref{eq:dpi}).

Let us confirm that a part of equations of motion matches with the equation
\ba
  E_{AB} \equiv R_{AB} - \frac{1}{2} \eta_{AB} R - \frac{1}{\ell^2} \eta_{AB} 
  + \beta D_C R_{D(A} \e_{B)}{}^{CD} = 0,
\end{alignat}
which is directly derived by the Lagrangian formalism in three dimensions.
Note that the indices $A,B,\cdots$ are used for three 
dimensional local Lorentz frame.
The covariant derivative $D_C$, which is defined by a spin connection, acts on the local Lorentz indices.
By contracting (\ref{eq:dotPi}) with $K_{ij}$ and using (\ref{eq:piplusA}) and (\ref{eq:Pi}), 
it is possible to derive
\ba
  0 &= - \beta \e^{ik} K_{ij} \dot{(g^{jl} K_{kl})}
  + 2 N K_{ij} K^{ij} - 2 N K^2 + 2\beta \e^{ik} K_{ij} \D_k \D^j N \notag
  \\
  &\quad\,
  - N \Big(r + \frac{2}{\ell^2} + K^{kl}K_{kl} - K^2 + 2\beta \e^{mn} \D_k \D_n K_m{}^k \Big) \notag
  \\
  &\quad\,
  - 2\beta \e^{mn} K_{ij} N^i \D_n K_m{}^j 
  + 2\beta \e^{ik} K_{ij} \D_k (N^l K_l{}^j) 
  - \beta \e^{ik} K_{ij} \D_l N^j K_k{}^l  \notag
  \\[0.1cm]
  &\quad\,
  + \beta \e^{ik} K_{ij} \D^j N^l K_{lk}
  - \beta \e^{ik} K_{ij} g^{jl} \dot K_{kl} \notag
  \\
  &= - N \Big(r + \frac{2}{\ell^2} - K^{kl}K_{kl} + K^2 \Big) 
  + 2\beta \e^{ij} \Big( \dot K_{ik} K_j{}^k + K_{ik} \D_j \D^k N \notag
  \\
  &\quad\,
  - N \D_k \D_j K_i{}^k 
  - N^l K_{kl} \D_j K_i{}^k + K_{ik} \D_j (N^l K_l{}^k) + \D^k N^l K_{ik} K_{jl} \Big).
\end{alignat}
After tedious calculations, we see that the expression is equal to $- 2 N E_{00}$, where
$0$ represents time direction in local Lorentz frame.
This gives a consistency check that we are dealing with the correct Hamiltonian.

As mentioned before, the variations of the Hamiltonian (\ref{eq:vH1}) and (\ref{eq:vH2}) are derived up
to total derivative terms. Therefore in order to derive correct equations of motion, it is necessary to
add surface term $Q[\xi]$ to the Hamiltonian. By taking the two 
 dimensional coordinates as $(r,\phi)$, 
the variation of the surface term $\dl Q[\xi]$ is given so as to cancel the total derivative terms
in the variation of the Hamiltonian:
\ba
  &\dl Q[\xi] \notag
  \\
  &= \int d\phi \Big[
  \sqrt{g} S^{ijkr} \big(\xi^0 \mc{D}_k \dl g_{ij} - \mc{D}_k \xi^0 \dl g_{ij} \big)
  + \xi^i (2 \pi^{jr} + \beta g^{\frac{1}{2}} A^{jr}) \dl g_{ij} 
  - \frac{1}{2} \xi^r (2 \pi^{ij} + \beta g^{\frac{1}{2}} A^{ij}) \dl g_{ij} \notag
  \\
  &\quad\,
  + \xi_i \dl (2 \pi^{ir} + \beta g^{\frac{1}{2}} A^{ir}) 
  - 2 \beta \sqrt{g} \e^{mr} \D_k \xi^0 g^{kl} \dl K_{ml}
  + 2 \beta \sqrt{g} \e^{mn} \xi^0 \D_n (g^{rl} \dl K_{ml}) \notag
  \\
  &\quad\,
  + \frac{1}{2} \beta \sqrt{g} S^{ijkr} \big( (\e^{mn} \pa_m \xi_n) \D_k \dl g_{ij}
  - \D_k (\e^{mn} \pa_m \xi_n) \dl g_{ij} \big) \notag
  \\
  &\quad\,
  - 2 \beta \sqrt{g} \e^{mr} \xi^i K_i{}^l \dl K_{ml}
  - \beta \sqrt{g} \e^{mn} \xi^i (\dl K_{ni}K_m{}^r + K_{ni} \dl K_m{}^r)
  - 2 \beta \sqrt{g} \e^{mn} g^{rl} \xi^0 K_{mo} \dl \gm^o{}_{nl} \notag
  \\[0.1cm]
  &\quad\,
  - 2 \beta \sqrt{g} \e^{mn} g^{kl} g^{op} \big\{ - \D_k \xi^0 K_{mo} T^{ijr}_{nlp}
  + \xi^0 \D_o K_{ml} T^{ijr}_{knp} + 2 \xi^0 \D_n K_{o(l} T^{ijr}_{m)kp} \big\} \dl g_{ij} \label{eq:TMGcharge}
  \\[0.1cm]
  &\quad\,
  + 2 \beta \sqrt{g} \e^{mn} \xi^o \big\{ K_o{}^p K_{mk} g^{kl} T^{ijr}_{npl} 
  + g^{qk} g^{lp} ( K_{mk} K_{l(o} T^{ijr}_{n)qp} 
  + K_{no} K_{l(k} T^{ijr}_{m)qp} ) \big\} \dl g_{ij} \notag
  \\
  &\quad\,
  + \frac{1}{2} \beta \sqrt{g} \e^{mn} T^{xyk}_{mlo} g^{op} \big\{ \D_k u_{xy} g^{lq} T^{ijr}_{npq}
  - u_{xy} \g^q{}_{np} g^{ls} T^{ijr}_{kqs} + 2 u_{xy} \g^l{}_{q(p} g^{qs} T^{ijr}_{n)ks} \big\} \dl g_{ij} \notag
  \\
  &\quad\,
  - \frac{1}{2} \beta \sqrt{g} \e^{mn} T^{ijr}_{mlo} u_{ij} g^{op} \dl \g^l{}_{np}
  + \frac{1}{2} \beta \sqrt{g} \e^{mr} \xi_m \dl r  \Big]. \notag
\end{alignat}
The index $r$ which is not contracted represents the radial coordinate.
Here we introduced $u_{ij}(\xi) \equiv 2 \xi^0 K_{ij} + 2 \D_{(i} \xi_{j)}$ 
and the vector $\xi=(\xi^0, \xi^r,\xi^\phi)$ is related to a Killing vector 
$\bar{\xi}=(\bar{\xi}^t, \bar{\xi}^r,\bar{\xi}^\phi)$ by (\ref{eq:localcoordinate}).
Since we are dealing with the Hamiltonian, the Killing vector should be $\bar{\xi}=(1,0,0)$ and $\xi=(N,N^r,N^\phi)$. 
When we deal with the angular momentum, the Killing vector should be $\bar{\xi}=(0,0,1)$ and $\xi=(0,0,1)$. 
We also call $\xi$ the Killing vector, since it does not make any confusion.
The equation (\ref{eq:TMGcharge}) makes it possible to evaluate the mass and the angular momentum of 
the BTZ black hole, and the central charges in TMG.


The charge $Q[\xi]$ itself is obtained by integrating (\ref{eq:TMGcharge}) over the canonical variables
with reference to the background values $\tilde{g}_{ij}$,
$\tilde{\pi}_{ij}$, $\tilde{K}_{ij}$ and $(2 \tilde{\pi}^{ir} + \beta \tilde{g}^{\frac{1}{2}} \tilde{A}^{ir})$. 
The last two become zero at the boundary.
(Consult Appendix \ref{App2} for explicit representations.) 
Then we are able to give an explicit form for $Q[\xi]$.
\ba
  Q[\xi] 
  &= \int d\phi \Big[
  \sqrt{\tilde{g}} \tilde{S}^{ijkr} \big(\xi^0 \tilde{\mc{D}}_k (g_{ij}-\tilde{g}_{ij}) 
  - \tilde{\mc{D}}_k \xi^0 (g_{ij}-\tilde{g}_{ij}) \big) \notag
  \\
  &\quad\,
  + \xi^i (2 \tilde{\pi}^{jr} + \beta \tilde{g}^{\frac{1}{2}} \tilde{A}^{jr}) (g_{ij}-\tilde{g}_{ij}) \notag
  - \frac{1}{2} \xi^r (2 \tilde{\pi}^{ij} + \beta \tilde{g}^{\frac{1}{2}} \tilde{A}^{ij}) (g_{ij}-\tilde{g}_{ij}) \notag
  \\
  &\quad\,
  + \xi_i (2 \pi^{ir} + \beta g^{\frac{1}{2}} A^{ir} ) 
  - 2 \beta \sqrt{\tilde{g}} \tilde{\e}^{mr} \tilde{\D}_k \xi^0 \tilde{g}^{kl} K_{ml} 
  + 2 \beta \sqrt{\tilde{g}} \tilde{\e}^{mn} \xi^0 \tilde{\D}_n (\tilde{g}^{rl} K_{ml}) \notag
  \\
  &\quad\,
  + \frac{1}{2} \beta \sqrt{\tilde{g}} \tilde{S}^{ijkr} \big( (\tilde{\e}^{mn} \pa_m \xi_n) 
  \tilde{\D}_k (g_{ij} - \tilde{g}_{ij})
  - \tilde{\D}_k (\tilde{\e}^{mn} \pa_m \xi_n) (g_{ij} - \tilde{g}_{ij}) \big) \Big]. \label{eq:integrated}
\end{alignat}
The canonical variables in $(\xi^0, \xi^i)$ should also be replaced by the background values.
Thus the integrability condition for $Q[\xi]$ is satisfied and $\delta
Q[\xi]$ is $\delta$-exact.
Note that in order to get this expression, we made use of $\tilde{K}_{ij} \to 0 \, (r \to \infty)$ and so on, 
so the terms in the last five lines in eq.~(\ref{eq:TMGcharge}) are simply dropped.
Note also that the integration constant in eq.~(\ref{eq:integrated}) is adjusted so that
the charge is zero for $(g_{ij},\pi_{ij},K_{ij},2 \pi^{ir} + \beta g^{\frac{1}{2}} A^{ir})
=(\tilde{g}_{ij},\tilde{\pi}_{ij},0,0)$. Explicit calculations of several charges are done
in the following subsection.

\subsection{Mass and Angular Momentum of BTZ Black Hole and Central Charges of CFT at the Boundary}

First let us evaluate the mass of the BTZ black hole in TMG by using (\ref{eq:TMGcharge}).
The BTZ black hole geometry (\ref{BTZsol})
still becomes a solution in TMG . 
This solution is invariant under the time translation and
corresponding Killing vector $\xi$ near the boundary is written as
\ba
  &(\xi^0, \xi^r, \xi^\phi) = (N, N^r, N^\phi) \sim \Big( \frac{r}{\ell},\; 0,\; \frac{4G_N j}{r^2} \Big).
\end{alignat}
In the background of BTZ black hole, two  dimensional quantities which are needed to estimate the mass
behave near the boundary as
\ba
  \dl g_{rr} \sim \frac{8G_N m\ell^4}{r^4}, \quad \dl K_{r\phi} \sim \frac{4G_N j\ell}{r^2}, \quad
  \dl (\pi^{r\phi} + \tfrac{1}{2} g^{\frac{1}{2}}A^{r\phi}) \sim \frac{4G_N j}{r^2}.
\end{alignat}
It seems that there are many terms to be estimated in (\ref{eq:TMGcharge}). 
Only a few terms, however, turn out to be non zero values.
In fact, last six lines in (\ref{eq:TMGcharge}) should be zero, since $\dl r \sim 0$, $\e^{mn}\pa_m N_n=0$, 
$K_{ij} = 0$ and $u_{ij}(\xi)=\dot g_{ij}=0$.
Some of remaining terms also vanish and the mass is eventually calculated as
\ba
  M &= \frac{1}{16\pi G_N} \dl Q[\xi] \notag
  \\
  &=\frac{1}{16\pi G_N} \oint_{r=\infty} d\phi
  \Big\{ 2 \sqrt{g} S^{r\phi r\phi} (- \xi^0 \gm^r{}_{\phi\phi} \dl g_{rr} )
  + 2 \beta \D_k \xi^0 g^{kl} \dl K_{\phi l} \Big\} \notag
  \\
  &= m + \frac{\beta}{\ell^2} j. \label{eq:mass}
\end{alignat}
For further details of the calculation the reader is referred to 
Appendix \ref{App2}.
This correctly reproduces the result obtained by the other methods~\cite{MCL,GHHM,DT,DKT,OST,kraus1,Solod,BC}.
The mass of the BTZ black hole in TMG 
is shifted by the angular momentum which is defined
in the gravity theory with negative cosmological constant.

The angular momentum of the BTZ black hole is also calculated in a similar way.
The BTZ solution is invariant under the rotation along the $\phi$ direction and corresponding Killing vector $\xi$ is written as
\ba
  &(\xi^0, \xi^r, \xi^\phi) = (0, 0, 1).
\end{alignat}
Last five lines in (\ref{eq:TMGcharge}) 
should be zero, since $\dl r \sim 0$, 
$K_{ij} = 0$ and $u_{ij}(\xi)=2\D_{(i} \xi_{j)}=0$.
The evaluation of remaining terms is not so laborious and the result becomes
\ba
  J &= \frac{1}{16\pi G_N} \dl Q[\xi] \notag
  \\
  &=\frac{1}{16\pi G_N} \oint_{r=\infty} d\phi
  \Big\{ \xi_i \dl (2\pi^{ir} + \beta g^{\frac{1}{2}}A^{ir})
  + \beta \sqrt{g} S^{r\phi r\phi} (- \e^{mn}\pa_m \xi_n \gm^r{}_{\phi\phi} \dl g_{rr} ) \Big\} \notag
  \\[0.2cm]
  &= j + \beta m. \label{eq:angular}
\end{alignat}
Again, the angular momentum of the BTZ black hole in TMG  is shifted by 
the mass which is defined in the gravity theory with negative 
cosmological constant~\cite{MCL,GHHM,DT,DKT,OST,kraus1,Solod,BC}.

Finally let us evaluate the central charge in TMG .
As discussed in Sec.  \ref{Sec3}, the diffeomorphisms which 
do not alter the boundary
condition are labelled by $\xi ^\pm_n$ in (\ref{eq:killing2})  
whose components are given by (\ref{eq:Killing}).
By using those Killing vectors, we can deform the global ${\rm AdS}_3$ 
background $G_{IJ}^0$. 
Here we choose a Killing vector $\bar{\eta}$ which corresponds 
to one of $\xi ^\pm_n$. Then the metric is written as
\ba
  G_{IJ} = G_{IJ}^0 + \D_I \bar{\eta}_J + \D_J \bar{\eta}_I.
\end{alignat}
From the ADM decomposition of this metric, the lapse $N$, the shift vector $N^i$ and two  dimensional metric $g_{ij}$
are obtained, and further it is possible to calculate the extrinsic curvature 
$K_{ij}\sim -\frac{1}{2N}(\D_i N_j + \D_j N_i)$ or canonical variables such as $\pi^{ij}$.
When we evaluate the central charge, we need to substitute these quantities into (\ref{eq:TMGcharge}).
Fortunately, since $\dl_\eta r$, $K_{ij}$ and $u_{ij}(\xi)$ are zero at the leading order, 
the expression for the central charge is simplified as
\ba
  &\dl_{\eta} Q[\xi] \notag
  \\
  &= \int d\phi \Big[
  \sqrt{g} S^{ijkr} \big(\xi^0 \mc{D}_k \dl_\eta g_{ij} - \mc{D}_k \xi^0 \dl_\eta g_{ij} \big)
  + \xi^i (2 \pi^{jr} + \beta g^{\frac{1}{2}} A^{jr}) \dl_\eta g_{ij} \notag
  \\[-0.1cm]
  &\qquad\qquad
  - \frac{1}{2} \xi^r (2 \pi^{ij} + \beta g^{\frac{1}{2}} A^{ij}) \dl_\eta g_{ij}
  + \xi_i \dl_\eta (2 \pi^{ir} + \beta g^{\frac{1}{2}} A^{ir})  \notag
  \\
  &\qquad\qquad
  - 2 \beta \sqrt{g} \e^{mr} \D_k \xi^0 g^{kl} \dl_\eta K_{ml}
  + 2 \beta \sqrt{g} \e^{mn} \xi^0 \D_n (g^{rl} \dl_\eta K_{ml}) \notag
  \\
  &\qquad\qquad
  + \frac{1}{2} \beta \sqrt{g} S^{ijkr} \big( (\e^{mn} \pa_m \xi_n) \D_k \dl_\eta g_{ij}
  - \D_k (\e^{mn} \pa_m \xi_n) \dl_\eta g_{ij} \big) \Big] \notag
  \\[0.1cm]
  &= \int d\phi \Big[
  \sqrt{g} S^{ijkr} \big(\xi^0 \mc{D}_k \dl_\eta g_{ij} - \mc{D}_k \xi^0 \dl_\eta g_{ij} \big)
  + \xi_i \dl_\eta (2 \pi^{ir} + \beta g^{\frac{1}{2}} A^{ir}) \label{eq:TMGcentralcharge}
  \\[-0.1cm]
  &\quad\,
  + 2 \beta \sqrt{g} \e^{rm} \pa_k \xi^0 g^{kl} \dl_\eta K_{ml} 
  + \frac{1}{2} \beta \sqrt{g} S^{ijkr} \big( (\e^{mn} \pa_m \xi_n) \D_k \dl_\eta g_{ij}
  - \D_k (\e^{mn} \pa_m \xi_n) \dl_\eta g_{ij} \big) \Big]. \notag
\end{alignat}
In the above, the Killing vector $\xi$ is constructed out of $\bar{\xi}$ as before, and its asymptotic value
should be chosen out of
\ba
  &(\xi^0, \xi^r, \xi^\phi) \sim \Big( \frac{r}{2} e^{inx^\pm}, -i \frac{nr}{2} e^{inx^\pm}, \pm \frac{1}{2} e^{inx^\pm} \Big).
\end{alignat}
Second equality is derived by substituting fluctuations of 
canonical variables. Those are summarized in  Appendix \ref{App2}.

Now we have prepared all tools which are necessary to 
calculate the central charge.
The central charge for $\eta=\xi ^+_n$ and $\xi=\xi ^+_m$ is evaluated as
\ba
  &\frac{1}{16\pi G_N} \dl_{\eta=\xi ^+_n} Q[\xi=\xi ^+_m] \notag
  \\
  &= \frac{1}{16\pi G_N} \oint_{r=\infty} d\phi \Big\{ \tf{1}{\ell} 
  \big( \tf{1}{r} \xi^0 + \pa_r \xi^0 \big) \dl_\eta g_{\phi\phi} 
  + \tf{r^3}{\ell^3} \xi^0 \dl_\eta g_{rr} 
  + \xi_\phi \dl_\eta (2 \pi^{\phi r} + \beta g^{\frac{1}{2}} A^{\phi r}) \Big\} \notag
  \\
  &\quad\,
  + \frac{\beta}{16\pi G_N} \oint_{r=\infty} d\phi \, \Big\{ 
  \tf{1}{\ell^2} \big( \tf{1}{r} \xi^0 + \pa_r \xi^0 \big) \dl_{\eta} g_{\phi\phi} 
  + \tf{r^3}{\ell^4} \xi^0 \dl_{\eta} g_{rr} + 2\pa_r \xi^0 g^{rl} \dl_{\eta} K_{\phi l} \Big\} \notag
  \\
  &= - \frac{i}{12}\frac{3\ell}{2G_N}\Big(1 + \frac{\beta}{\ell} \Big) m(m^2-1) \dl_{m,-n}.
\end{alignat}
The second line just gives the contribution of Einstein-Hilbert 
term.  The third line gives the modification.
In a similar way the central charge for $\eta=\xi ^-_n$ 
and $\xi=\xi ^-_m$ is evaluated as
\ba
  &\frac{1}{16\pi G_N} \dl_{\eta=\xi ^-_n} Q[\xi=\xi ^-_m] \notag
  \\
  &= \frac{1}{16\pi G_N} \oint_{r=\infty} d\phi \Big\{ \tf{1}{\ell} 
  \big( \tf{1}{r} \xi^0 + \pa_r \xi^0 \big) \dl_\eta g_{\phi\phi} 
  + \tf{r^3}{\ell^3} \xi^0 \dl_\eta g_{rr} 
  + \xi_\phi \dl_\eta ( 2 \pi^{\phi r} + \beta g^{\frac{1}{2}} A^{\phi r}) \Big\} \notag
  \\
  &\quad\,
  + \frac{\beta}{16\pi G_N} \oint_{r=\infty} d\phi \, \Big\{ 
  - \tf{1}{\ell^2} \big( \tf{1}{r} \xi^0 + \pa_r \xi^0 \big) \dl_{\eta} g_{\phi\phi} 
  - \tf{r^3}{\ell^4} \xi^0 \dl_{\eta} g_{rr} + 2 \pa_r \xi^0 g^{rl} \dl_{\eta} K_{\phi l}\Big\} \notag
  \\
  &= - \frac{i}{12}\frac{3\ell}{2G_N}\Big(1 - \frac{\beta}{\ell} \Big) m(m^2-1) \dl_{m,-n}. 
\end{alignat}
Again the central charge is given by the sum of the Einstein-Hilbert 
part and the gravitational Chern-Simons part.
Note, however, that the signs in front of the modifications are 
different.   From a similar calculation, it is possible to check 
that $\dl_{\eta=\xi ^+_n} Q[\xi=\xi ^-_m] = 0$.
If we call $\xi ^+_m$ left mover and $\xi ^-_m$ right one, the central 
charges are written as~\cite{kraus2,kraus1,sahoo,Solod}
\ba
  c_L &= \frac{3 \ell}{2G_N} \Big(1 + \frac{\beta}{\ell} \Big), \notag
  \\
  c_R &= \frac{3 \ell}{2G_N} \Big(1 - \frac{\beta}{\ell} \Big).
  \label{eq:virasorcentralcharges}
\end{alignat}
Thus via canonical formalism of topologically massive gravity, we have succeeded to realize 
the Virasoro algebras of left and right movers with different central 
charges\footnote{See ref.~\cite{bc} in the case of Riemann-Cartan geometry.}.

\section{Central Charges with All Higher Derivative Corrections}\label{Sec5}

\subsection{Final Expression of Central Charges}
\label{susec:finalexpression}

We have by now established the canonical formalism  and have 
got the Virasoro central charges (\ref{eq:virasorcentralcharges}) 
for TMG  which consists of Einstein-Hilbert action with negative cosmological 
constant and gravitational Chern-Simons terms. 
We are now in a position to generalize these results in order to encompass 
most general cases of higher derivative gravity.  
In the last paragraph of Sec.~\ref{Sec3},  
we have seen  that inclusion of  all higher derivative 
corrections other than  the gravitational Chern-Simons term requires us 
to multiply the central charges of Brown and Henneaux's 
 by the conformal factor $\Omega$ \cite{saida}.
Making use of this simple scaling rule, we get to the 
following final expression of the central charges for the 
left and right movers:
\ba
  c_L &= \frac{3 \ell}{2G_N} \Big(\Omega + \frac{\beta}{\ell} \Big), \notag
  \\
  c_R &= \frac{3 \ell}{2G_N} \Big(\Omega - \frac{\beta}{\ell} \Big).
\label{finalc}
\end{alignat}
We would like to emphasize that all of the effects due to higher order terms 
are included in the factors $\Omega$ and $\beta$. 
We also note that 
these central charges are obtained by  constructing 
Virasoro algebras directly in the canonical method without referring to 
the  Wald's formula or variations thereof.

Furthermore, the definition of mass and 
angular momentum of BTZ black hole should necessarily be modified  
in the most general theory of gravity (\ref{generalaction}). 
Combining the results of (\ref{eq:massang}), (\ref{eq:mass}) and (\ref{eq:angular}),
the effective mass and the angular momentum become
\begin{eqnarray}
M=\Omega m+\frac{\beta }{\ell ^{2}}j, 
\hskip0.5cm 
J=\Omega j +\beta m . 
\label{newMJ}
\end{eqnarray}
From these, the zero modes $L_0^+$ and $L_0^-$ of Virasoro algebras for left 
and right movers are expressed as
\begin{eqnarray}
L_{0}^{+}=\frac{1}{2} \left ( M \ell + J \right ) 
=\left(\Omega  + \frac{\beta}{\ell} \right ) \frac{1}{2}  \left ( m \ell + j\right),
\\
L_{0}^{-}=\frac{1}{2} \left ( M \ell - J \right ) 
=\left(\Omega  - \frac{\beta}{\ell} \right ) \frac{1}{2}  \left ( m \ell - j\right).
\end{eqnarray}
%
Putting these together with (\ref{finalc}) into the 
 Cardy's formula for counting the states in CFT, we obtain the entropy 
\begin{eqnarray}
  S &=& 2\pi \sqrt{\frac{1}{6}c_{L} L_{0}^{+}} + 
  2\pi \sqrt{\frac{1}{6}c_{R} L_{0}^{-} } \notag
  \\
  &=& \frac{\pi}{2 G_N} \left  ( \Omega + \frac{\beta}{\ell} \right ) 
  \sqrt{2 G_{N} \ell ^{2} \left ( m +  \frac{j}{\ell } \right )} 
  +\frac{\pi}{2 G_N} \left  ( \Omega - \frac{\beta}{\ell} \right ) 
  \sqrt{2 G_{N} \ell ^{2} \left ( m -  \frac{j}{\ell } \right )} . 
  \label{eq:finalentropy}
\end{eqnarray}
This agrees  with the previous entropy formula (\ref{macroentropy}) 
obtained by the extended Wald's formula.
For the BTZ black hole capturing the contributions of all higher 
derivative corrections, we have thus proved the agreement between 
the macroscopic entropy and the Cardy's entropy of microstate counting.


\subsection{Realization in M-theory: M5 System}

In \ref{susec:finalexpression}, we applied Brown-Henneaux's method to the 
three dimensional gravity theories
with most general higher derivative terms, and derived the central charges 
of CFT at the boundary.  
The three dimensional theory of that sort is 
usually embedded in higher dimensional 
theories in the string theory context. 
Among several others, the most interesting  example is 
embodied  in M-theory, which is intriguing because 
  the corresponding CFT is understood very clearly~\cite{msw}.

The M-theory is defined in eleven dimensions and its low energy limit 
is well described by eleven dimensional supergravity.
When we compactify the eleven dimensional supergravity on ${\rm CY}_3$, 
it becomes five dimensional supergravity with eight supercharges.
Beyond the low energy limit, the M-theory contains a lot of terms which 
correct eleven  dimensional supergravity.  Among subleading terms 
in the derivative expansion, there 
exists a would-be Chern-Simons term which is expressed as~\cite{DLM}
\ba
  \frac{\ell_p^6}{2\kappa_{11}^2} \frac{\pi^2}{3 \cdot 2^6} 
  \int A \wedge \text{tr} (R \wedge R) \wedge \text{tr} (R \wedge R),
\end{alignat}
where $\ell_p$ is the Planck length in eleven dimensions, $2\kappa_{11}^2 = (2\pi)^8 \ell_p^9$.
$A$ is a 3-form potential and the M5-brane is magnetically coupled to this field.  
The Chern-Simons term in five dimensions arises after reducing this term. 
In fact, by expanding the 3-form with a basis of harmonic $(1,1)$-forms $J_{\hat I}$ in CY$_3$ as
$A= 8\pi^2 \ell_p^3 A^{\hat I} \wedge J_{\hat I}$, we obtain~\cite{AFMN}
\ba
  \frac{c_{2 \hat I} }{3\cdot 2^7 \pi^2} \int A^{\hat{I}}\wedge \text{tr}(R \wedge R). \label{eq:5dCS}
\end{alignat}
Here $A^{\hat I}$ corresponds to gauge fields in five dimensions,
and $c_{2 \hat I} = \frac{1}{8 \pi^2} \int_{\text{CY}_3} J_{\hat I} \wedge \text{tr}(R \wedge R)$.
In general, other curvature squared terms are also obtained by the dimensional 
reduction of $R^4$ terms in the M-theory.
It is, however, complicated to work out all these terms, thus we employ the five dimensional
conformal supergravity below.

By applying  the conformal supergravity approach, the action of the five dimensional supergravity 
theory concerned with $R^2$ terms has been  constructed in ref.~\cite{hot1}.
With this action together with an ansatz of AdS$_3$ $\times$ S$^2$ 
geometry, we would like to  identify  the effective 
cosmological constant $-2/ \ell^2$, 
the conformal factor $\Omega$ and the coupling constant $\beta$ in 
terms of topological quantities in CY$_3$.
Actually the three dimensional gravitational Chern-Simons term can be obtained 
by compactifying (\ref{eq:5dCS}) on S$^2$.
The central charges are also expressed by the CY$_{3}$ data 
in the context of string theory.
In passing note that similar calculations have been  done 
in refs.~\cite{kraus2,kraus1, gupta}.

In the following we employ notations used in ref.~\cite{gupta}.
Assuming that the five dimensional metric, gauge fields and 2-form auxiliary field $v$ be given by 
\begin{align}
ds^2_{(5)}&=\psi^2G_{IJ}dx^Idx^J+\chi^2d\Omega_{S^2}^2,\notag\\
F^{\hat{I}}_{\theta\phi}&=\frac{p^{\hat{I}}}{2}\sin\theta,\notag\\
v_{\theta\phi}&=V\sin\theta,
\end{align}
we obtain the three dimensional supergravity with the curvature 
squared terms and gravitational Chern-Simons term. Here 
$p^{\hat{I}}$ corresponds to the M5-brane charge.
In order to realize the Einstein frame in three dimensions, we have to set
\begin{equation}
\psi^{-1}=\frac{\chi^2}{\pi}\left(
\frac{3}{4}+\frac{1}{4}\mathcal{N}
+\frac{c_{2{\hat{I}}}M^{\hat{I}}}{288\chi^2}
+\frac{c_{2{\hat{I}}}M^{\hat{I}}V^2}{72\chi^4}
-\frac{c_{2{\hat{I}}}p^{\hat{I}}V}{288\chi^4}
\right),
\end{equation}
where $M^{\hat I}$ stand for moduli scalars. 
Then the action becomes \cite{gupta}
\begin{align}
\mathcal{S}=\int d^3x \sqrt{-G}\left(
R+Z(\phi)+A(\phi)R^2+B(\phi)R_{IJ}R^{IJ}
\right)
+\int d^3x\mathcal{L}_{\text{CS}}
+\mathcal{S}',
\label{m5action}
\end{align}
in the unit of $16\pi G_N=1$.
Here $\phi$ stands generically for all scalars $M^{\hat{I}}\,,\,V\,,\,D$ and $\chi$, 
and $\mathcal{S}'$ includes their derivative terms.
The scalar potential and each coupling are given by
\begin{align}
Z(\phi)&=\psi^3\frac{\chi^2}{\pi}\Bigg\{
\frac{2}{\chi^2}\left(
\frac{3}{4}+\frac{\mathcal{N}}{4}
\right)
-2\left(
\frac{D}{4}-\frac{V^2}{\chi^4}
\right)
+\mathcal{N}\left(
\frac{D}{2}+\frac{6V^2}{\chi^4}
\right)
+\frac{2\mathcal{N}_{\hat{I}}p^{\hat{I}}V}{\chi^4}
+\frac{\mathcal{N}_{{\hat{I}}{\hat{J}}}p^{\hat{I}}p^{\hat{J}}}{8\chi^4}
\notag\\
&+\frac{c_{2{\hat{I}}}M^{\hat{I}}}{96\chi^4}
+\frac{c_{2{\hat{I}}}M^{\hat{I}}D^2}{288}
+\frac{c_{2{\hat{I}}}p^{\hat{I}}VD}{144\chi^4}
-\frac{5c_{2{\hat{I}}}M^{\hat{I}}V^2}{36\chi^6}
-\frac{c_{2{\hat{I}}}p^{\hat{I}}V}{48\chi^6}
+\frac{c_{2{\hat{I}}}p^{\hat{I}}V^3}{36\chi^8}
+\frac{c_{2{\hat{I}}}M^{\hat{I}}V^4}{6\chi^8}
\Bigg\},\notag
\end{align}
\begin{equation}
A(\phi)=-\frac{5}{6}\frac{c_{2{\hat{I}}}M^{\hat{I}}\chi^2}{192\pi \psi}
\,\,\,\,,\,\,\,\,
B(\phi)=\frac{8}{3}\frac{c_{2{\hat{I}}}M^{\hat{I}}\chi^2}{192\pi \psi}
\,\,\,\,,\,\,\,\,
\beta=-\frac{c_{2{\hat{I}}}p^{\hat{I}}}{96\pi},
\end{equation}
where
\begin{equation}
\mathcal{N}=\frac{1}{6}c_{{\hat{I}}{\hat{J}}{\hat{K}}}M^{\hat{I}}M^{\hat{J}}M^{\hat{K}}\,\,\,\,,\,\,\,\,
\mathcal{N}_{\hat{I}}=\frac{1}{2}c_{{\hat{I}}{\hat{J}}{\hat{K}}}M^{\hat{J}}M^{\hat{K}}\,\,\,\,,\,\,\,\,
\mathcal{N}_{{\hat{I}}{\hat{J}}}=c_{\hat{I}\hat{J}\hat{K}}M^{\hat{K}}.
\end{equation}
$c_{{\hat{I}}{\hat{J}}{\hat{K}}}$ and $c_{2{\hat{I}}}$ are the 
triple intersection number and the second Chern class number of 
${\rm CY}_3$, respectively.

The action (\ref{m5action}) enables us to derive an equation of motion for the metric
\begin{align}
\frac{1}{2}G^{IJ}\{
R+AR^2+B(R_{IJ})^2+Z
\}
&-R^{IJ}-2ARR^{IJ}-2BR^{IK}R^J_K+T^{IJ}\notag\\
&=\beta \epsilon^{KL(I}\mathcal{D}_KR^{J)}_L
+(\textrm{derivative terms of }\phi),
\label{m5einstein}
\end{align}
and those for scalars
\begin{equation}
\partial_\phi Z+\partial_\phi AR^2
+\partial_\phi B(R_{IJ})^2
=(\textrm{derivative terms of }\phi).
\label{m5scalar}
\end{equation}
It is, however, almost impossible for us to find general solutions 
to these equations.  What we can do is  to take  all 
$\phi$ to be constants everywhere and to assume that the BTZ black 
hole which satisfies (\ref{ads3sol}).
It corresponds to the black ring solution whose geometry is 
AdS$_3$ $\times$ S$^2$ in five dimension.
By substituting (\ref{ads3sol}), equations of motion (\ref{m5einstein}) and (\ref{m5scalar}) reduce to
\begin{align}
Z&=\frac{2}{\ell^2}+(3A+B)\left(\frac{2}{\ell^2}\right)^2,\notag\\
\partial_\phi Z&+3\,(3\,\partial_\phi A+\partial_\phi B)\left(\frac{2}{\ell^2}\right)^2
=0.
\label{m5eom}
\end{align}
Since $c_{2\hat{I}}$ indicates the higher derivative corrections in the next order, 
we can solve five equations (\ref{m5eom}) to the first order of $c_{2\hat{I}}$.
The solutions are
\begin{align}
M^{\hat{I}}=\frac{p^{\hat{I}}}{p}\left(1-\frac{C}{36}\right)\,\,,\,\,
V=-\frac{3}{8}p\left(1+\frac{C}{36}\right)\,\,,\,\,
D=\frac{12}{p^2}\left(1-\frac{C}{18}\right)\,\,,\,\,
\chi=\frac{p}{2}\left(1+\frac{C}{36}\right),
\end{align}
and
\begin{align}
\ell=\frac{p^3}{4\pi}\left(1+\frac{37}{288}C\right),
\end{align}
where
$p^3=\frac{1}{6}c_{{\hat{I}}{\hat{J}}{\hat{K}}}p^{\hat{I}}p^{\hat{J}}p^{\hat{K}}$ and $C=c_{2\hat{I}}p^{\hat{I}}/p^3$.
On the other hand, the conformal factor $\Omega$ for this solution is calculated as 
\begin{align}
\Omega(\ell)&=1+2AR+\frac{2}{3}BR\notag\\
&\simeq1-\frac{C}{288}.
\end{align}

The assumption of constant scalars admits  the BTZ black hole solution.
Therefore, the Brown-Henneaux's approach explained in the previous sections can be applied to this solution, and we can prove the existence 
of  the two dimensional CFT satisfying the Virasoro algebra on the 
AdS boundary.
With the use of $16\pi G_N=1$, $\beta=-c_{2\hat{I}}p^{\hat{I}}/96\pi$ 
and the formula (\ref{finalc}), the central charges of the left and 
right movers are given by
\begin{alignat}{3}
c_L&=6p^3+\frac{1}{2}c_{2{\hat{I}}}p^{\hat{I}}, \notag
\\
c_R&=6p^3+c_{2{\hat{I}}}p^{\hat{I}},
\end{alignat}
in agreement with \cite{msw, kraus2, kraus1, cdkl1, gupta}. 
In four dimensions, these central charges appear in expressions of the 
entropy for the extremal non-BPS and BPS black holes, respectively \cite{dst}. 
The precise information of the microstates for the CFT at the boundary 
is veiled  in our formalism.  Whatever the microstates may be, 
we can only see the Virasoro algebras and calculate their central charges.
But as for the M5-brane system, the explanation for microstates was made 
clear in ref.~\cite{msw}  from the detailed description of the effective 
field theory on the brane.



\section{Summary and Discussions}

In the present paper we have analyzed topologically massive gravity (\ref{eq:action}), 
 using the conventional canonical formalism. Since there are 
higher derivative terms w.r.t.  time, we made use of the generalized version of the
Ostrogradsky method. 
We defined the global charges so as to cancel the surface terms of the variation of the Hamiltonian.
Using these, we have derived the Virasoro algebras
realized asymptotically at the boundary ($r \sim \infty$), and found that the
central charges are given by (\ref{eq:virasorcentralcharges}), which do not respect left-right symmetry. 
The mass and the angular momentum of the BTZ black hole are also
computed including the effects due to the gravitational Chern-Simons term. 

We have gone one step further to 
argue that effects due to higher derivative terms  can be  included
by employing the scaling argument. 
The central charges, the mass and the 
angular momentum in such a general class of higher derivative theories 
are given by (\ref{finalc}) and (\ref{newMJ}), respectively. 
The BTZ black hole entropy is given by (\ref{eq:finalentropy}), 
which agrees with the formula given by using the modified Wald's 
formula.  We have thus succeeded in strengthening the link between 
the two dimensional boundary CFT  and the three dimensional 
gravity  description of the black hole. 
As an interesting example, we considered the three dimensional model which are realized 
by compactifying the M-theory on CY$_3$ $\times$ S$^2$. The left-right asymmetric central charges are 
already given in ref.~\cite{msw} from the microscopic viewpoint, and we confirmed that the result
can exactly be reproduced in our formalism.
The consideration from the representation of Virasoro algebras is also important as future 
works~\cite{NY}.

In this paper, we treated $\ell$ and $\beta$ as free parameters.
From the analyses by CFT, however, it was pointed out that the three dimensional gravity theory 
could be realized so as to be consistent with unitarity and positivity
only when $\ell$ and $\beta$ take some particular values~\cite{witten2,lss}\footnote{For recent
arguments, see refs.~\cite{CDWW,GJ,LSS2}.}.
It would be interesting if we could re-examine these observations from our canonical approach
including the higher derivative contribution $\Omega$.
Generalizations of our formalism to the supergravity or inclusion of other matter fields are also interesting future directions~\cite{deser:sugra,nos}.

\vspace{0.2cm}
\noindent
{\bf Acknowledgements}

KH is supported in part by JSPS Research Fellowship for Young Scientists. 
The work of YH is partially supported by the Ministry of Education, Science, 
Sports and Culture, Grant-in-Aid for Young Scientists (B), 19740141, 2007.

\appendix
\setcounter{equation}{0}
\setcounter{figure}{0}
\renewcommand{\theequation}{A.\arabic{equation}}
\renewcommand{\thefigure}{A.\arabic{figure}}


\section{ADM Decomposition of Gravitational Chern-Simons Term}\label{App1}

In this Appendix  we give a quick summary of  the ADM decomposition 
of the gravitational Chern-Simons term.
First note that capital variables $G_{IJ}$, $E^A_{\,\,\,I}$, 
$\Gamma^I{}_{JK}$ and $\Omega^A_{\,\,\,BI}$ are three dimensional 
metric, vielbein, affine connection and spin connection, respectively, 
and  $g_{ij}$, $e^a_{\,\,\,i}$, $\gamma^i{}_{jk}$ and $\omega^a_{\,\,\,bi}$ 
are two dimensional ones. As in the main body of the text, 
$I,J\cdots=t,r,\phi$ labels the three dimensional space-time indices 
and $A,B\cdots=0,1,2$ denotes the three dimensional local Lorentz indices.
Also, small indices are of two dimensional: $i,j\cdots=r,\phi$ and 
$a,b\cdots=1,2$.

For the purpose of the  ADM decomposion of the gravitational 
Chern-Simons term (\ref{eq:csaction}), the simplest way 
is  to divide ${\cal L}_{{\rm CS}}$ into the Lorentz Chern-Simons 
and remaining terms.
When we write the connection 1-form expressed by the matrix notation as $\Gamma^I_{\,\,\,J}=\Gamma^I{}_{JK}dx^K$, the relation between the affine connection and the spin connection is
\begin{equation}
\Gamma^I_{\,\,\,J}=E^I_{\,\,\,A}\Omega^A_{\,\,\,B}E^B_{\,\,\,J}+E^I_{\,\,\,A}dE^A_{\,\,\,J}.
\end{equation}
Note that $\Gamma^I_{\,\,\,J}$ and $\Omega^A_{\,\,\,B}$ are the connection 1-form but $E^A_{\,\,\,J}$ is the 0-form vielbein.
Omitting the indices like
$\Gamma=E^{-1}\Omega E+E^{-1}dE$,
the gravitational Chern-Simons term is expressed by 
\begin{equation}
\textrm{Tr}\left(
\Gamma d\Gamma+\frac{2}{3}\Gamma^3
\right)
=\textrm{Tr}\left(
\Omega d\Omega+\frac{2}{3}\Omega^3
\right)
-\frac{1}{3}\textrm{Tr}(dEE^{-1})^3
-d\,\textrm{Tr}(dEE^{-1}\Omega).
\label{gammaomega}
\end{equation}
We can drop the last term since it is just a total derivative.

Let us define 
\begin{equation}
K_{ab}=\frac{1}{N}(e^i_{\,\,(\,a}\dot{e}_{b\,)\,i}-D_{(\,a}N_{b\,)})\,\,\,\,\,\,,\,\,\,\,\,\,
L_{ab}=\frac{1}{N}(e^i_{\,\,[\,a}\dot{e}_{b\,]\,i}-D_{[\,a}N_{b\,]}),
\end{equation}
where $D_i$ is the covariant derivative which acts on the 
local Lorentz indices like $D_iV^a=\partial_iV^a+\omega^a_{\,\,\,bi}V^b$.
Then the first term of (\ref{gammaomega}) becomes
\begin{eqnarray}
\Omega^A_{\,\,\,B} d\Omega^B_{\,\,\,A}
&=&2\Omega^0_{\,\,\,a} d\Omega^a_{\,\,\,0}
+\Omega^a_{\,\,\,b} d\Omega^b_{\,\,\,a}\notag\\
&=& 2\{
-(\partial_a N+K_{ab}N^b)\partial_jK^a_{\,\,\,i}
+\partial_j(\partial^a N+K^a_{\,\,\,b}N^b)K_{ai}
+K_{aj}\dot{K}^a_{\,\,\,i}
\}dt\wedge dx^i\wedge dx^j\notag\\
& &+\{
\omega^a_{\,\,\,bi}\partial_j(L^b_{\,\,\,a}N)
-\partial_j\omega^b_{\,\,\,ai}L^a_{\,\,\,b}N
-\omega^a_{\,\,\,bi}\dot{\omega}^b_{\,\,\,aj}
\}dt\wedge dx^i\wedge dx^j,
\end{eqnarray}
and the second term is
\begin{eqnarray}
\Omega^A_{\,\,\,B}\Omega^B_{\,\,\,C}\Omega^C_{\,\,\,A}
&=&3\Omega^0_{\,\,\,a}\Omega^a_{\,\,\,b}\Omega^b_{\,\,\,0}\notag\\
&=&3(
2\partial_aN\omega^a_{\,\,\,bi}K^b_{\,\,\,j}
-NK_{ai}K^b_{\,\,\,j}L^a_{\,\,\,b}
+K_{aj}K^b_{\,\,\,i}\omega^a_{\,\,\,bc}N^c
\notag \\
& & 
+2K_{ae}K^b_{\,\,\,j}\omega^a_{\,\,\,bi}N^e
)dt\wedge dx^i\wedge dx^j.
\end{eqnarray}
If we use $\epsilon^{tij}\sqrt{-G}=\epsilon^{ij}\sqrt{g}$ and the two dimensional identity 
$r^{ab}_{\,\,\,\,\,ij}=re^{[\,a}_{\,\,\,[\,i}e^{b\,]}_{\,\,\,j\,]}$,
the Lorentz Chern-Simons term can be written as
\begin{eqnarray}
\textrm{Tr}\left(
\Omega d\Omega+\frac{2}{3}\Omega^3
\right)
& \cong & 
\{
4(\partial_a N+K_{ab}N^b)D_{i}K^a_{\,\,\,j}
+2K_{aj}\dot{K}^a_{\,\,\,i}-\omega^a_{\,\,\,bi}
\dot{\omega}^b_{\,\,\,aj}\notag\\
& &\,\,\,\,\,\,+(2K_{aj}K^b_{\,\,\,i}+re^b_{\,\,\,i}e_{aj})(D_bN^a-e^k_{\,\,\,b}\dot{e}^a_{\,\,\,k})
\}\sqrt{g}\epsilon^{ij}d^3x\notag\\
&=&\{
4(\partial_l N+K_{lk}N^k)\mathcal{D}_{i}K^l_{\,\,\,j}
+2\dot{K}^k_{\,\,\,i}K_{kj}
+2K_{kj}K^l_{\,\,\,i}\mathcal{D}_{l}N^k
+r\mathcal{D}_{i}N_j\notag\\
& &\,\,\,\,\,\,-re_{aj}\dot{e}^a_{\,\,\,i}
-\omega^a_{\,\,\,bi}\dot{\omega}^b_{\,\,\,aj}
\}\sqrt{g}\epsilon^{ij}d^3x,
\label{lorentz}
\end{eqnarray}
up to total derivative.
$K_{ij}$ is of course the extrinsic curvature (\ref{eq:kij})
and $\mathcal{D}_i$ is the usual covariant derivative.

On the other hand, one can check that the second  term 
on the right hand side of (\ref{gammaomega})  becomes
\begin{eqnarray}
-\frac{1}{3}\textrm{Tr}(dEE^{-1})^3
&=&-\frac{1}{3}(dEE^{-1})^a_{\,\,\,b}(dEE^{-1})^b_{\,\,\,c}(dEE^{-1})^c_{\,\,\,a}\notag\\
&=&\dot{e}^k_{\,\,\,b}\partial_ie^b_{\,\,\,l}e^l_{\,\,\,c}
\partial_je^c_{\,\,\,k}dt\wedge dx^i\wedge dx^j.
\label{eq:a7}
\end{eqnarray}
After some manipulation  and neglecting the total derivative, 
the sum of the last lines in (\ref{lorentz}) and (\ref{eq:a7}) becomes 
\begin{align}
&\hskip-2cm \int d^3x\sqrt{g}\epsilon^{ij}
(
-re_{aj}\dot{e}^a_{\,\,\,i}
-\omega^a_{\,\,\,bi}\dot{\omega}^b_{\,\,\,aj}
+\dot{e}^k_{\,\,\,b}\partial_ie^b_{\,\,\,l}e^l_{\,\,\,c}
\partial_je^c_{\,\,\,k}
)\notag\\
=&
  \int d^3x \sqrt{g}  \e^{ij} \dot \g^l{}_{ik} \g^k{}_{jl} \notag\\
=& \int d^3x \sqrt{g}(-\e^{mp} g^{lo} T^{ijk}_{mno} \D_k \g^n{}_{pl} )\dot{g}_{ij},
\label{gammagamma}
\end{align}
in which we used the definition (\ref{eq:tijkmno}) of $T^{ijk}_{mno}$  
and 
\begin{eqnarray}
\D_k \g^n{}_{pl}\equiv \partial_k\g^n{}_{pl}+\g^n{}_{km}\g^m{}_{pl}
-\g^m{}_{kp}\g^n{}_{ml}-\g^m{}_{kl}\g^n{}_{pm}. 
\end{eqnarray}
Defining $A^{ij}$ by (\ref{eq:aij}), 
we can rewrite the last line of (\ref{gammagamma}) as
\begin{eqnarray}
-\int d^3x \sqrt{g}A^{ij}\dot{g}_{ij}
&=&-\int d^3x \sqrt{g}A^{ij}(2NK_{ij}+2\D_{i} N_{j})
\notag\\
 & \cong & \int d^3x \sqrt{g}(-2A^{ij}NK_{ij}+2\D_{i}A^{ij}N_{j}).
\end{eqnarray}
In fact, the result of the central charge for the Virasoro algebra 
does not depend  on whether we use the gravitational Chern-Simons 
or Lorentz Chern-Simons.  As we have seen in Sec.~\ref{Sec4}, 
this is due to the fact 
that the above terms involving $A^{ij}$ have no contribution to 
the central charge calculation.
To sum up, combination of  all the terms calculated above leads us 
finally to the following expression:
\ba
  &\sqrt{-G} \epsilon^{IJK} \Big( \Gamma^P{}_{IQ} \pa_J \Gamma^Q{}_{KP}
  + \frac{2}{3} \Gamma^P{}_{IQ} \Gamma^Q{}_{JR} \Gamma^R{}_{KP} \Big) \notag
  \\
  &=\sqrt{g}\big[
 2  \e^{mn} \dot K_{mk} K_n{}^k 
  +  N \big\{ 4 \e^{mn} \D_k \D_n K_m{}^k - 2 A^{kl}K_{kl} \big\} \notag
  \\
  &\quad\,
  +  N^i \big\{ - 4 \e^{mn} K_i{}^l \D_n K_{ml} - 2 \e^{mn} \D_k(K_{ni}K_m{}^k)
  + g_{im} \e^{mn} \pa_n r + 2 \D_k A_i{}^k \big\}
\big]. 
\end{alignat}

\renewcommand{\theequation}{B.\arabic{equation}}
\renewcommand{\thefigure}{B.\arabic{figure}}


\section{Supplementary Calculations on Charges}\label{App2}

\subsection{Mass and angular momentum of BTZ black hole}

Let us consider the ADM decomposition of the BTZ black hole solution (\ref{BTZsol}).
From the canonical procedure, the lapse, the shift vector and the two dimensional metric are given by
\ba
  &N =\tilde{N} + \dl N \sim \frac{r}{\ell} - \frac{4 G_N m \ell}{r}, \notag
  \\
  &N^r =\tilde{N}^r + \dl N^r \sim 0 + 0, \notag
  \\
  &N^\phi =\tilde{N}^\phi + \dl N^\phi \sim 0 + \frac{4 G_N j}{r^2}, 
  \\
  &g_{ij} =\tilde{g}_{ij} + \dl g_{ij} \sim 
  \begin{pmatrix}
    \frac{\displaystyle{\ell^2}}{\displaystyle{r^2}} & 0 \\
    0 & r^2
  \end{pmatrix} +
  \begin{pmatrix}
    \frac{\displaystyle{8 G_N m \ell^4}}{\displaystyle{r^4}} & 0 \\
    0 & 0
  \end{pmatrix}. \notag
\end{alignat}
These are expanded around $m=j=0$ and the flucutuations are linearly dependent on $m$ or $j$.
From the metric $\tilde{g}_{ij}$ given in the above, nonzero components of the affine connection and 
$\tilde{S}^{ijkl} = \tf{1}{2} (\tilde{g}^{ik}\tilde{g}^{jl} + \tilde{g}^{il}\tilde{g}^{jk} - 2\tilde{g}^{ij}\tilde{g}^{kl})$ are evaluated as
\ba
  &\tilde{\gm}^r{}_{rr} = -\frac{1}{r}, \quad \tilde{\gm}^\phi{}_{r\phi} = \frac{1}{r}, \quad
  \tilde{\gm}^r{}_{\phi\phi} = -\frac{r^3}{\ell^2}, \notag
  \\
  &\tilde{S}^{\phi\phi rr} = -\frac{1}{\ell^2}, \quad \tilde{S}^{\phi r\phi r} = \frac{1}{2\ell^2}. \label{eq:S}
\end{alignat}
By using the equations of motion (\ref{eq:g}) and (\ref{eq:dpi}), the extrinsic curvature
and the covariant combination of the momentum are calculated as
\ba
  &K_{ij} =\tilde{K}_{ij} + \dl K_{ij} \sim
  \begin{pmatrix}
    0 & 0 \\
    0 & 0
  \end{pmatrix} +
  \begin{pmatrix}
    0 & \frac{\displaystyle{4 G_N j \ell}}{\displaystyle{r^2}} \\
    \frac{\displaystyle{4 G_N j \ell}}{\displaystyle{r^2}} & 0
  \end{pmatrix}, \notag
  \\
  &(\pi+\tfrac{1}{2}\beta g^{\frac{1}{2}} A)^{ij} =(\tilde{\pi}+\tfrac{1}{2}\beta \tilde{g}^{\frac{1}{2}} \tilde{A})^{ij}+ \dl (\pi+\tfrac{1}{2}\beta g^{\frac{1}{2}} A)^{ij} \sim
  \begin{pmatrix}
    0 & 0 \\
    0 & 0
  \end{pmatrix} +
  \begin{pmatrix}
    0 & \frac{\displaystyle{4 G_N j}}{\displaystyle{r^2}} \\
    \frac{\displaystyle{4 G_N j}}{\displaystyle{r^2}} & 0
  \end{pmatrix}.
\end{alignat}
In order to derive these expressions, we dropped terms with time derivatives.

\subsection{The central charges}

Let us consider the geometry constructed by $G_{IJ}=G_{IJ}^0 + \D_I \bar \eta_J + \D_J \bar \eta_I$.
Here $G_{IJ}^0$ is the metric of global AdS$_3$ and $\bar \eta_I$ represents some Killing vector.
For $\eta = \xi _n^\pm$, the lapse, the shift vector and the two dimensional metric behave asymptotically as
\ba
  &N=\tilde{N} + \dl N \sim \Big(\frac{r}{\ell} + \frac{\ell}{2r} \Big) - \frac{i \ell n (n^2-2)}{4r} e^{inx^\pm}, \notag
  \\
  &N^r =\tilde{N}^r + \dl N^r \sim 0 - \frac{\ell n^2}{r} e^{inx^\pm}, \notag
  \\
  &N^\phi =\tilde{N}^\phi + \dl N^\phi \sim 0 \pm \frac{i \ell n (n^2-1)}{2r^2} e^{inx^\pm}, 
  \\
  &g_{ij} =\tilde{g}_{ij} + \dl_\eta g_{ij} \sim 
  \begin{pmatrix}
    \frac{\displaystyle{\ell^2}}{\displaystyle{r^2}} - 
    \frac{\displaystyle{\ell^4}}{\displaystyle{r^4}} & 0 \\
    0 & r^2
  \end{pmatrix} +
  \begin{pmatrix}
    - \frac{\displaystyle{i n \ell^4}}{\displaystyle{r^4}} & 
    \mp \frac{\displaystyle{n^2 \ell^4}}{\displaystyle{2 r^3}} 
    \\[0.2cm]
    \mp \frac{\displaystyle{n^2 \ell^4}}{\displaystyle{2 r^3}} & 
    \frac{\displaystyle{i n^3 \ell^2}}{\displaystyle{2}} 
  \end{pmatrix} e^{inx^\pm}. \notag
\end{alignat}
From the metric $\tilde{g}_{ij}$ given in the above, asymptotic behaviors of the affine connection and 
$\tilde{S}^{ijkl}$ are evaluated as (\ref{eq:S}).
By using the equations of motion (\ref{eq:g}) and (\ref{eq:dpi}), the extrinsic curvature
and the covariant combination of the momentum are calculated as
\ba
  &K_{ij} =\tilde{K}_{ij} + \dl_\eta K_{ij} \sim
  \begin{pmatrix}
    0 & 0 \\
    0 & 0
  \end{pmatrix} +
  \begin{pmatrix}
    -\frac{\displaystyle{2 n^2 \ell^4}}{\displaystyle{r^5}} & 
    \pm \frac{\displaystyle{i n(n^2-1) \ell^2}}{\displaystyle{2r^2}} 
    \\[0.2cm]
    \pm \frac{\displaystyle{i n(n^2-1) \ell^2}}{\displaystyle{2r^2}} &
     \frac{\displaystyle{n^2(n^2+1) \ell^2}}{\displaystyle{2r}}
  \end{pmatrix} e^{inx^\pm} ,\notag
  \\[0.1cm]
  &(\pi+\tfrac{1}{2}\beta g^{\frac{1}{2}} A)^{ij} =(\tilde{\pi}+\tfrac{1}{2}\beta \tilde{g}^{\frac{1}{2}} \tilde{A})^{ij}+ 
  \dl_\eta (\pi+\tfrac{1}{2}\beta g^{\frac{1}{2}} A)^{ij} \notag
  \\
  &\quad\, \sim
  \begin{pmatrix}
    0 & 0 \\
    0 & 0
  \end{pmatrix} +
  \begin{pmatrix}
    -\frac{\displaystyle{n^2 (\ell\mp\beta+n^2(\ell\pm\beta))}}
    {\displaystyle{2r}} & \pm \frac{\displaystyle{i n(n^2-1) \ell}}
    {\displaystyle{2r^2}} 
    \\[0.2cm]
    \pm \frac{\displaystyle{i n(n^2-1) \ell}}{\displaystyle{2r^2}} 
    & 
    \pm\frac{\displaystyle{n^2(\pm 4\ell+\beta(n^2-1)) \ell^2}}
    {\displaystyle{2r^5}}
  \end{pmatrix} e^{inx^\pm} .
\end{alignat}
It is useful to note that $\tilde{\D}_n (\tilde{g}^{rl} \dl_\eta K_{ml})$ is 
symmetric under the exchange of $m$ and $n$, i.e., 
\ba
  \tilde{\D}_n (\tilde{g}^{rl} \dl_\eta K_{ml}) &=
  \begin{pmatrix}
    \frac{\displaystyle{6 n^2 \ell^2}}{\displaystyle{r^4}} & 
    \mp \frac{\displaystyle{i n(n^2-1)}}{\displaystyle{r}} 
    \\[0.2cm]
    \mp \frac{\displaystyle{i n(n^2-1)}}{\displaystyle{r}} & 
    -n^2(n^2+2)
  \end{pmatrix} e^{inx^\pm} .
\end{alignat}
Other useful equations used in the text are : 
\ba
  &\tilde{g}^{rl} \dl_\eta K_{\phi l} \sim \pm i \tf{1}{2}n(n^2 - 1) e^{inx^\pm}, \quad
  \e^{pq} \pa_p \xi_q \sim \pm \frac{r}{\ell} e^{inx^\pm} \sim \pm \frac{2}{\ell} \xi^0.
\end{alignat}

 

\begin{thebibliography}{99} 
%
%
\bibitem{deser}
S. Deser and R. Jackiw, Ann. Phys. {\bf 153} (1984) 405. 

\bibitem{btz}
M. Ba${\tilde {\rm n}}$ados, C. Teitelboim and J. Zanelli, 
Phys. Rev. Letters. {\bf 69} (1992) 1849, hep-th/9204099. 

\bibitem{witten1}
A. Ach\'ucarro and P.K. Townsend, Phys. Lett. {\bf B180} (1986) 89. \\
E. Witten, Nucl. Phys. {\bf B311} (1988) 46.
%
\bibitem{brown}
J.D. Brown and M. Henneaux, Commun. Math. Phys. {\bf 104} (1986) 207.

\bibitem{msw}
J.M. Maldacena, A. Strominger and E. Witten, JHEP {\bf 9712} (1997) 002, hep-th/9711053.

\bibitem{CCAF}
A.C. Cadavid, A. Ceresole, R. D'Auria and S. Ferrara, Phys. Lett. {\bf B357} (1995) 76, hep-th/9506144.

\bibitem{kraus2}
P. Kraus and F. Larsen, JHEP {\bf 0509} (2005) 034, hep-th/0506176.

\bibitem{kraus1}
P. Kraus and F. Larsen, JHEP {\bf 0601} (2006) 022, hep-th/0508218.

\bibitem{Krauslec}
As a recent review see for example, P. Kraus, hep-th/0609074.
%
\bibitem{saida}
H. Saida and J. Soda, Phys. Lett. {\bf B471} (2000) 358, gr-qc/9909061.

\bibitem{MFF}
G. Magnano, M. Ferraris and M. Francaviglia, Gen. Rel. Grav. {\bf 19} (1987) 465.
%
\bibitem{gupta}
R.K. Gupta and A. Sen, JHEP {\bf 0803} (2008) 015, arXiv:0710.4177[hep-th]. 


\bibitem{Deser:1982vy}
  S.~Deser, R.~Jackiw and S.~Templeton,
  Phys.\ Rev.\ Letters  {\bf 48} (1982) 975.

\bibitem{Deser:1981wh}
  S.~Deser, R.~Jackiw and S.~Templeton,
  Ann. Phys.\  {\bf 140} (1982) 372
  [Erratum-ibid.\  {\bf 185} (1988) 406];
  Ann. Phys.\  {\bf 281} (2000) 409.




\bibitem{ost}
M.V. Ostrogradsky, Mem. Acad. Sci., St. Petersberg {\bf 6} (1850) 385.
%
\bibitem{buchbinder}
I.L. Buchbinder and S.L. Lyahovich, Class. Quant. Grav. {\bf 4} (1987) 1487.
%
\bibitem{buchbinder2}
I.L. Buchbinder, S.L. Lyakhovich and V.A. Krykhtin, Class. Quant. Grav. {\bf 10} (1993) 2083.
%
\bibitem{sahoo}
B. Sahoo and A. Sen, JHEP {\bf 0607} (2006) 008, hep-th/0601228.

\bibitem{Solod}
S.N. Solodukhin, Phys. Rev. {\bf D74} (2006) 024015, hep-th/0509148.

\bibitem{wald}
R.M. Wald, Phys. Rev. {\bf D48} (1993) R3427, gr-qc/9307038;
 
V. Iyer and R.M. Wald, Phys. Rev. {\bf D50} (1994) 846, gr-qc/9403028.

\bibitem{s}
Y. Tachikawa, Class. Quant. Grav. {\bf 24} (2007) 737, hep-th/0611141.

%
%
%
%
\bibitem{p}
M.-I. Park, Phys. Rev. {\bf D77} (2008) 026011, hep-th/0608165; Phys. Rev. {\bf D77} (2008) 126012, hep-th/0609027; 
Class. Quant. Grav. {\bf 25} (2008) 135003, arXiv:0705.4381[hep-th].
%



%
\bibitem{RT}
T. Regge and C. Teitelboim, Ann. Phys. {\bf 88} (1974) 286.
%
%




\bibitem{Deser:1991qk}
  S.~Deser and X.~Xiang,
  Phys.\ Lett.\  {\bf B263} (1991) 39.


\bibitem{MCL}
K.A. Moussa, G. Clement and C. Leygnac, Class. Quant. Grav. {\bf 20} (2003) L277, gr-qc/0303042.

\bibitem{GHHM}
A.A. Garcia, F.W. Hehl, C. Heinicke and A. Macias, Phys. Rev. {\bf D67} (2003) 124016, gr-qc/0302097.

\bibitem{DT}
S. Deser and B. Tekin, Class. Quant. Grav. {\bf 20} (2003) L259, gr-qc/0307073.

\bibitem{DKT}
S. Deser, I. Kanik and B. Tekin, Class. Quant. Grav. {\bf 22} (2005) 3383, gr-qc/0506057.

\bibitem{OST}
S. Olmez, O. Sarioglu and B. Tekin, Class. Quant. Grav. {\bf 22} (2005) 4355, gr-qc/0507003.

\bibitem{BC}
A. Bouchareb and G. Clement,
Class. Quant. Grav. {\bf 24} (2007) 5581, arXiv:0706.0263 [gr-qc].



\bibitem{bc}
M. Blagojevic and B. Cvetkovic, 'Trends in General Relativity and Quantum Cosmology'. Volume 2. (2006) 85, gr-qc/0412134.

\bibitem{DLM}
M.J. Duff, J.T. Liu and R. Minasian, Nucl. Phys. {\bf B452} (1995) 261, hep-th/9506126.


\bibitem{AFMN}
I. Antoniadis, S. Ferrara, R. Minasian and K.S. Narain, Nucl. Phys. {\bf B507} (1997) 571, hep-th/9707013.

\bibitem{hot1}
K. Hanaki, K. Ohashi and Y. Tachikawa, Prog. Theor. Phys. {\bf 117} (2007) 533, hep-th/0611329.


\bibitem{cdkl1}
A. Castro, J.L. Davis, P. Kraus and F. Larsen, JHEP {\bf 0704} (2007) 091, hep-th/0702072.


\bibitem{dst}
A. Dabholkar, A. Sen and S.P. Trivedi, JHEP {\bf 0701} (2007) 096, hep-th/0611143.



\bibitem{NY}
N. Yokoi and T. Nakatsu, Prog. Theor. Phys. {\bf 104} (2000) 439, hep-th/9912096.



\bibitem{witten2}
E. Witten, arXiv:0706.3359 [hep-th]

\bibitem{lss}
W. Li, W. Song and  A. Strominger, JHEP {\bf 0804} (2008) 082, arXiv:0801.4566[hep-th].

\bibitem{deser:sugra}
S. Deser, ``Cosmological Topological Supergravity'' in Quantum Theory of Gravity, 
edited by S.M. Christensen, Adam Hilger, London, 1984.

\bibitem{nos}
M. Natsuume, T. Okamura and M. Sato,  Phys. Rev. {\bf D61} (2000) 104005, hep-th/9910105.

\bibitem{CDWW}
S. Carlip, S. Deser, A. Waldron and D.K. Wise,
arXiv:0803.3998 [hep-th].

\bibitem{GJ}
D. Grumiller and N. Johansson,
arXiv:0805.2610 [hep-th].

\bibitem{LSS2}
W. Li, W. Song and  A. Strominger,
arXiv:0805.3101 [hep-th].




\end{thebibliography}
\end{document}